\documentstyle[11pt,epsfig]{article}
\input{epsf}
\begin{document}
\setlength{\baselineskip}{0.18in}
\newcommand{\be}{\begin{eqnarray}}
\newcommand{\ee}{\end{eqnarray}}
\newcommand{\bi}{\bibitem}
\newcommand{\nue}{\nu_e}
\newcommand{\num}{\nu_\mu}
\newcommand{\nut}{\nu_\tau}
\newcommand{\nus}{\nu_s}
\newcommand{\nua}{\nu_a}
\newcommand{\mne}{m_{\nu_e}}
\newcommand{\mnm}{m_{\nu_\mu}}
\newcommand{\mnt}{m_{\nu_\tau}}
\newcommand{\lar}{\leftarrow}
\newcommand{\rar}{\rightarrow}
\newcommand{\lrar}{\leftrightarrow}
\newcommand{\nuh}{\nu_h}
\newcommand{\mnh}{m_{\nu_h}}
\newcommand{\taut}{\tau_{\nut}}
\newcommand{\fg}{f_{\gamma}}
\newcommand{\dE}{{\delta E}}
\newcommand{\dm}{{\delta m^2}}
\newcommand{\rss}{\rho_{ss}}
\newcommand{\ras}{\rho_{as}}
\newcommand{\rsa}{\rho_{sa}}
\newcommand{\rba}{\rho_{ba}}
\newcommand{\rab}{\rho_{ab}}
\newcommand{\raa}{\rho_{aa}}
\newcommand{\rcb}{\rho_{cb}}
\newcommand{\rac}{\rho_{ac}}
\newcommand{\rsb}{\rho_{sb}}
\newcommand{\rbs}{\rho_{bs}}
\newcommand{\hee}{{\cal H}_{ee}}
\newcommand{\hmm}{{\cal H}_{\mu\mu}}
\newcommand{\hem}{{\cal H}_{e\mu}}
\newcommand{\hes}{{\cal H}_{es}}
\newcommand{\hms}{{\cal H}_{\mu s}}
\newcommand{\hmt}{{\cal H}_{\mu \tau}}
\newcommand{\het}{{\cal H}_{e \tau}}
\newcommand{\hts}{{\cal H}_{\tau s}}

\newcommand{\dms}{\Delta_{\mu s}}
\newcommand{\dme}{\Delta_{\mu e}}
\newcommand{\dmt}{\Delta_{\mu \tau}}
\newcommand{\des}{\Delta_{e s}}

\newcommand{\deta}{\Delta_{e \tau}}
\newcommand{\dts}{\Delta_{\tau s}}
\newcommand{\dsm}{\Delta_{s\mu }}
\newcommand{\dem}{\Delta_{e\mu}}
\newcommand{\dtm}{\Delta_{\tau \mu}}
\newcommand{\dse}{\Delta_{se}}
\newcommand{\dte}{\Delta_{\tau e}}
\newcommand{\dst}{\Delta_{s\tau }}

\newcommand{\gee}{\gamma_{ee}}
\newcommand{\gem}{\gamma_{\mu\mu}}
\newcommand{\ges}{\gamma_{es}}
\newcommand{\gms}{\gamma_{\mu s}}
\newcommand{\rem}{\rho_{e\mu}}
\newcommand{\rme}{\rho_{\mu e}}
\newcommand{\res}{\rho_{es}}
\newcommand{\rse}{\rho_{se}}
\newcommand{\rms}{\rho_{\mu s}}
\newcommand{\rsm}{\rho_{s\mu}}
\newcommand{\rmm}{\rho_{\mu \mu}}
\newcommand{\ree}{\rho_{ee}}
\begin{center}
\vglue .06in
{\Large \bf { 
BBN bounds on active-sterile neutrino mixing
  }
}
\bigskip
\\{\bf A.D. Dolgov}$^{(a)(b)(c)(d)}$ and
{\bf F.L. Villante}$^{(a)(d)}$
\\[.2in]
$^{(a)}${\it INFN, sezione di Ferrara,
Via Paradiso, 12 - 44100 Ferrara,
Italy} \\
$^{(b)}${\it ITEP, Bol. Cheremushkinskaya 25, Moscow 113259, Russia
}  \\
$^{(c)}${\it Research Center for the Early Universe, Graduate School of 
Science, \\University of Tokyo, Tokyo 113-0033, Japan}\\
$^{(d)}${\it Dipartimento di Fisica, Universita' di Ferrara, 
Via Paradiso 12 - 44100 Ferrara, Italy}
\\

%\\{\bf A.D. Dolgov} \\
% \\[.5in]
%{\it{INFN, sezione di Ferrara,
%Via Paradiso, 12 - 44100 Ferrara,
%Italy \\
%and 
%ITEP, Bol. Cheremushkinskaya 25, Moscow 113259, Russia.
%}} \\[.2in]
%{\bf F. L. Villante} \\
% \\[.5in]
%{\it{
%INFN, sezione di Ferrara,
%Via Paradiso, 12 - 44100 Ferrara,
%Italy
%}}\\
\end{center}

\vspace{.3in}
\begin{abstract}
Nucleosynthesis restrictions on mixing of active neutrinos 
with possible sterile ones are obtained with the account of 
experimentally determined mixing between all active
neutrinos. The earlier derived bounds, valid in the absence of
active-active mixing, are reanalyzed and significant difference 
is found in the resonance case. The results are
obtained both analytically and numerically by solution of complete
system of integro-differential kinetic equations. 
A good agreement between analytical and numerical
approaches is demonstrated. A role of possibly large cosmological
lepton asymmetry
is discussed.
\end{abstract}

\section{Introduction \label{s-intr}}

Combined data from  SNO~\cite{sno}, KamLAND~\cite{kamland}, and
earlier data on solar~\cite{solar-nu} and atmospheric~\cite{atm-nu}
neutrino deficit present strong 
indication that all these neutrino anomalies can be explained without 
introduction of a new sterile neutrino, for recent analysis
see refs.~\cite{fogli02}-\cite{holanda02}).
According to ref.~\cite{bahcall02}, 
a mixing of active, $\nua$ ($a=e,\mu,\tau$), 
and sterile, $\nus$, neutrinos is excluded at the level $\sin \eta < 0.13$.
On the other hand, at a weaker level, mixing with $\nus$ is not only allowed
but even desirable. According the ref.~\cite{smirnov-ven}, some features of
the solar neutrino data are better described if there exists 
fourth sterile neutrino
state with the mixing parameter $\sin^2 2\theta = 10^{-4}-10^{-3}$ and 
the mass difference $\dm  =  (10^{-5}-10^{-6})$ eV$^2$. Moreover, the
LSND results~\cite{lsnd} probably also demands mixing with additional one or 
several  new (sterile) neutrinos or maybe, if confirmed, 
something even more drastic.

Consideration of big bang nucleosynthesis (BBN) may 
noticeably strengthen the upper bound on mixing between active and 
sterile neutrinos and even exclude explanation of LSND anomaly by 
admixture of $\nus$. Such limits can also restrict explanations of solar and
atmospheric neutrino features by an admixture of $\nus$. 
BBN bounds on mixing between $\nus$ and $\nua$ were derived long time 
ago~\cite{old-nus}-\cite{enqvist92} 
under assumption that only one active neutrino is mixed with a sterile
partner and mixing between active neutrinos was disregarded
(for more references and discussion see the review~\cite{dolgov02}). 
A simplified consideration of 4 neutrino mixing has 
been done recently in ref.~\cite{dibari03}
for rather large mixing angle, $\sin^2 2\theta_{as} >0.01$.

It is practically established now that all active neutrinos,
$\nue$, $\num$, and $\nut$, are strongly mixed with the parameters given 
by Large Mixing Angle (LMA) solution to solar neutrino 
deficit~\cite{holanda02}:
\be
\dm_{\rm sol} = 7.3\cdot 10^{-5}\,\,\,{\rm eV}^2, \,\,\,
\tan^2 \theta_{\rm sol} =0.4
\label{lma}
\ee
and by atmospheric neutrino data~\cite{atm-nu}:
\be
\dm_{\rm atmo} = 2.5\cdot 10^{-3}\,\,\,{\rm eV}^2, \,\,\,
\tan^2 \theta_{\rm atmo} \approx 1
\label{atm}
\ee
Existence of fast transitions between $\nue$, $\num$, and $\nut$ 
noticeably changes BBN bound on mixing with sterile neutrinos especially
for small values of mass difference. The reason for this change is the
following. There are three possible effects on BBN created by mixing 
between active and sterile neutrinos. First is the production of additional 
neutrino species in the primeval plasma. The second
effect is a depletion of the number density of electronic neutrinos
which results in a higher neutron freezing temperature. Both these
effects lead to a larger neutron-to-proton ratio and to 
more abundant production of primordial deuterium and helium-4 (for 
the details see e.g. review~\cite{dolgov02}).
If mixing between active neutrinos is absent then the second 
effect would manifest itself only in the case of $(\nue-\nus)$-mixing,
if we neglect relatively weak depopulation of $\nue$ through the annihilation
$\bar\nue \nue \rar \bar\nu_{\mu,\tau} \nu_{\mu,\tau}$. In the realistic 
case of experimentally measured mixing between all active neutrinos a 
deficit of $\num$ or $\nut$ would be efficiently transformed into 
deficit of $\nue$ leading to a stronger BBN bound on active-sterile 
mixing. 

The third effect is a generation of large lepton asymmetry due to 
oscillations between active and sterile species~\cite{res-rise}.
However, this effect takes place only for very weak mixing, much
smaller than the experimental bound and is neglected
throughout almost all the paper, except for sec.~\ref{s-leptas}.

The paper is organized as follows. In sec.~\ref{s-kineq} general
kinetic equations for oscillating neutrinos in the primeval cosmological
plasma are presented. In sec.~\ref{s-prel} 
the approximations that we use for analytical
and numerical calculations are discussed and the effects of neutrino on 
Big Band Nucleosynthesis (BBN)
are described. In sec.~\ref{s-one+one} we reconsider the case of 
mixing between
one active and one sterile neutrinos, assuming that other active neutrinos 
are not mixed. Our calculations agree with the earlier works in 
non-resonance case, 
while there is a noticeable difference in the resonance case. 
In. sec~\ref{s-two+one} we consider for illustrative 
purposes algebraically 
much simpler situation when only two active neutrinos are mixed.  
The realistic case of all three
active neutrinos mixed is presented in sec.~\ref{s-three+one}. 
In all previous sections the calculations are done for negligibly small 
cosmological lepton asymmetry. The impact of the latter is discussed in 
sec.~\ref{s-leptas}. In conclusion we summarize and discuss the 
derived bounds.

%%%%%%%%%%%%%%%%%%%
\section{Kinetic equations \label{s-kineq}}
%%%%%%%%%%%%%%%%%%%%%%%

We assume that in addition to three known active neutrinos there exists
a fourth neutrino flavor which does not have any interactions except for 
mixing with active neutrinos. The transformation between 
 flavor eigenstates $\nu_\alpha$ and mass eigenstates
$\nu_j$ is described by the orthogonal matrix
\be
\nu_\alpha = U_{\alpha j} \,\nu_j
\label{nuj-nua}
\ee
where $\alpha = e,\mu,\tau,s$ and $j=1,2,3,4$.
We disregard here CP-violation effects, so matrix $U$ is assumed to be 
real. In the limit of small mixing $\nue \approx \nu_1$, 
$\num \approx \nu_2$, $\nut\approx \nu_3$, and $\nus \approx \nu_4$.
We know, however, that the real mixing between active neutrinos is 
large and they cannot be taken as a dominant single mass eigenstate.

Kinetic equations for the density matrix of oscillating
neutrinos have the usual form~\cite{kin-eq}:
\be
i\dot \rho = [{\cal H}^{(1)},\rho ] - i \{{\cal H}^{(2)},\rho \} 
\label{dotrho}
\ee
where the first commutator term includes vacuum Hamiltonian and effective
potential of neutrinos in medium calculated in the first order
in Fermi coupling constant, $G_F$.
The effective potential is proportional to the deviation
of neutrino refraction index from unity and is calculated in ref.~\cite{nora}.
The second anti-commutator term
includes imaginary part of the Hamiltonian calculated in the second
order in $G_F$
\footnote{ Here and in the following we will use a natural system of
unit in which $\hbar=c=1$.}. 
This term describes breaking of coherence induced by
neutrino scattering and annihilation as well as neutrino production
by collisions in primeval plasma. 

Written in components kinetic equations have the form:
\be
i\dot \rss &=&  \sum_a {\cal H}^{(1)}_{sa} (\ras -\rsa )
\label{dot-rss} \\
i\dot \raa &=&  \sum_b {\cal H}^{(1)}_{ab} (\rba -\rab )
+ {\cal H}^{(1)}_{as} (\rsa - \ras) -i I_{coll} (\rho)
\label{dot-raa}\\
i\dot \rab &=&  {\cal H}^{(1)}_{as} \rsb - {\cal H}^{(1)}_{sb} \ras
+ \sum_c \left( {\cal H}^{(1)}_{ac} \rcb - {\cal H}^{(1)}_{cb}\rac \right)
-i \gamma_{ab} \rab 
\label{dot-rab}\\
i\dot \rsa &=&  {\cal H}^{(1)}_{ss} \rsa -  {\cal H}^{(1)}_{sa} \rss
+\sum_b \left(  {\cal H}^{(1)}_{sb}\rba -
{\cal H}^{(1)}_{ba}\rsb \right)- i\gamma_{as}\rsa
\label{dot-rsa}
\ee
where $a,b,c$ label active neutrino species. Matrix elements of
the first order Hamiltonian 
$ {\cal H}^{(1)}_{\alpha\beta}=  {\cal H}^{(1)}_{\beta\alpha} $ 
(below we will omit the upper index 1) are given by:
\be
{\cal H}_{\alpha \beta} = V_{\alpha \beta} + 
\sum_{j=1-4} (m_j^2/2E) U_{\alpha j} U_{\beta j}
\label{h-alphabeta}
\ee 
where $E$ is the neutrino energy and
$\alpha$ and $\beta$ run over all neutrino flavors, active and sterile.
The effective potential in the medium, $V_{\alpha \beta}$,
vanishes if any $\alpha$ or $\beta$ are equal to $s$. For active neutrinos
diagonal components of the potential are given by~\cite{nora}:
\be 
V_{aa} = \pm C_1 \eta^{(a)} G_FT^3 -
C_2^a \frac{G^2_F T^4 E}{\alpha}
\label{veff}
\ee
where $T$ is the plasma temperature, 
$G_F$ is the Fermi coupling
constant, $\alpha=1/137$ is the fine structure constant, and the signs
``$\pm$'' refer to neutrinos and anti-neutrinos respectively.
The first term arises due to charge asymmetry in the primeval plasma,
while the second one comes from non-locality of weak interactions 
associated with the exchange of $W$ or $Z$ bosons. 
According to ref.~\cite{nora} the
coefficients $C_j$ are: $C_1 \approx 0.95$, $C_2^e \approx 0.61$ and
$C_2^{\mu,\tau} \approx 0.17$ (for $T<m_\mu$).  
These values are true in the limit of
thermal equilibrium, otherwise these coefficients are some
integrals from the distribution functions over momenta.
The charge asymmetry of plasma is described by the coefficients
$\eta^{(a)}$ which are equal to
\be
\eta^{(e)}& =&
2\eta_{\nue} +\eta_{\num} + \eta_{\nut} +\eta_{e}-\eta_{n}/2 \,\,\,
 ( {\rm for} \,\, \nue)~,
\label{etanue} \\
\eta^{(\mu)} &=&
2\eta_{\num} +\eta_{\nue} + \eta_{\nut} - \eta_{n}/2\,\,\,
({\rm for} \,\, \num)~,
\label{etanumu}
\ee
and $\eta^{(\tau)}$ for $\nut$ is obtained from eq.~(\ref{etanumu}) by
the interchange $\mu \lrar \tau$. The individual charge asymmetries,
$\eta_X$, are defined as the ratio of the difference between
particle-antiparticle number densities to the 
number density of photons with the account of the 11/4-factor emerging
from $e^+e^-$-annihilation:
\be
\eta_X = \left(N_X -N_{\bar X}\right) /N_\gamma
\label{etax}
\ee
If $\nu\nu$-interactions are essential then off-diagonal components 
$V_{ab}$ are non-vanishing~\cite{off-diag} (see also recent discussion in
refs.~\cite{pastor01rs}-\cite{friedland03}). These components are 
proportional to the integrals over neutrino momenta from off-diagonal 
components of the density matrix. In the case under consideration
these non-diagonal components of the effective potential are sub-dominant
(see sec.~\ref{s-two+one}) and will be neglected.

The coherence breaking terms in off-diagonal components of the
density matrix are given by 
$\gamma_{\alpha\beta} = (\gamma_\alpha + \gamma_\beta )/2$, where
$\gamma_s =0$ and $\gamma_a$ is the total reaction rate, including
elastic scattering and annihilation of $\nu_a$: 
\be 
\gamma_a = g_a \frac{180 \zeta(3)}{7 \pi ^4} \, G_F^2 T^4 p ~.
\label{gamma-a}
\ee
Here $p$ is the neutrino momentum and the coefficients $g_a$,
according to ref.~\cite{enqvist92} are $g_{\nu_e} \simeq 4$ and 
$g_{\nu_\mu,\mu_\tau} \simeq 2.9$. Slightly smaller results
$g_{\nu_e} \simeq 3.56$ and $g_{\nu_\mu,\mu_\tau} \simeq 2.5$
were obtained in ref.~\cite{dolgov00} due to account of Fermi
statistics.

The coherence breaking terms in kinetic equations for diagonal
components of density matrix, $\raa$, are given by
\be
I_{coll} =&&{1\over 2E_1} \sum \int  d\tau (l_2,\nu_3,l_4)
|A_{el}|^2 \left[ \raa (p_1) f_l(p_2) - \raa (p_3) f_l(p_4)\right]-
\nonumber \\
&&{1 \over 2E_1} \sum \int d\tau (l_2,\nu_3,l_4)
|A_{ann}| \left[\raa(p_1) \bar\raa (p_2)-f_l(p_3)f_{\bar l}(p_4)
\right]
\label{I-coll}
\ee
where $A_{el,ann}$ are the amplitudes of the corresponding reactions
and the sum is taken over all possible channels of elastic scattering
(first term) and annihilation (second term). The phase space element
is given by:
\be
d\tau (l_2,\nu_3,l_4) = (2\pi)^4 \delta^4 (p_1+p_2-p_3-p_4)\,
{d^3 p_2 \over 2E_2 \, (2\pi)^3}\,{d^3 p_3 \over 2E_3 \, (2\pi)^3}\,
{d^3 p_4 \over 2E_4 \, (2\pi)^3}
\label{dtau}
\ee

It is noteworthy that in the absence of annihilation 
$\bar\nua \nua \lrar e^+e^-$ the total number density of all neutrinos is
conserved in comoving volume:
\be
a^3(t) \int {d^3 p \over (2\pi)^3} \left(\rss +\sum_a \raa\right) 
= const
\label{n-tot}
\ee
where $a(t)$ is the cosmological scale factor. At the temperatures
below freezing of the reactions $\nua \bar \nua \lrar e^+ e^-$ 
($T^f_{\nue} \approx 3.2 $ MeV and $T^f_{\num,\nut} \approx 5.3 $ 
MeV~\cite{dolgov02,dolgov-itep}, see also sec.~\ref{s-kin})
sterile neutrino states are produced at the expense of active ones and
the effective number of additional neutrino species does not change.
However the effect of the oscillations on BBN would be
noticeable even in this case because of the deficit of $\nue$ induced 
either by 
direct mixing $\nue - \nus$ or by mixing of $\nu_{\mu,\tau} - \nus$ and 
fast redistribution of active neutrino species due to large mixing 
between them.

Concluding this section, we introduce convenient variables in terms
of which we will solve kinetic equations:
\be 
x = m a(t) = m/T,\,\,\, y=E a(t) = E/T
\label{xy}
\ee
where $m$ is an arbitrary normalization mass which for convenience
is taken as 1 MeV and $a(t)$ is the cosmological scale factor. We
assume that the temperature evolves as $T= 1/a(t)$. Though it is not
always true, the  precision of this assumption is sufficiently good for 
our purposes. In terms of these variables the l.h.s. of kinetic
equations in Friedman-Robertson-Walker background can be rewritten as
\be
\dot f = \left( \partial_t - Hp \partial_p\right) f = 
Hx \partial_x f
\label{lhs}
\ee
where $H=\dot a /a$ is the Hubble parameter. 

Expressed through $x$ and $y$, essential quantities take the following forms:
\be 
H = h/x^2, \,\,\, {\rm with} \,\,\, 
h = 4.46\cdot 10^{-22} \left( g_* /10.75 \right)^{1/2}
\label{H}
\ee
where $g_*$ is the number of species in the primeval plasma;  $g_* = 10.75$ for 
$T<m_\mu$; we have neglected here presumably small ($<1$) contribution from $\nus$. 
The energy difference of neutrinos is given by
\be
\dE \equiv \frac{\dm}{2E}
= 5\cdot 10^{-13} \left( \dm /{\rm eV}^2 \right) (x/y) \equiv  d_m(x/y)
\label{dE}
\ee 
Below we will always measure neutrino mass difference in eV$^2$ if
not stated otherwise.
The ``non-local'' contribution to neutrino potential in matter
(the second term in eq.~(\ref{veff})) is equal to
\be
V^{(a)}_{nl} = C^{(a)}_v (y/x^5)
\label{vnl}
\ee
where $ C^{(e)}_v = 1.137\cdot 10^{-20}$ and
$ C^{(\mu)}_v = C^{(\tau)}_v =  0.317\cdot 10^{-20}$. 
Contribution to the potential from charge asymmetry of primeval plasma 
is always sub-dominant if $\eta$ has ``normal value $10^{-9} - 10^{-10}$. 
The case of non-negligible charge asymmetry is considered in 
sec.~\ref{s-leptas}. The coherence 
breaking terms determined by $\gamma$ have the same dependence on $x$
and $y$ as $V^{(a)}_{nl}$ but with much smaller coefficient. Still they
should be retained because only they contribute into imaginary parts
of the density matrix and create exponential damping. They are given by:
\be
\gamma_{ab} = \epsilon_{ab} (y/x^5)
\label{gammaab}
\ee
where $\epsilon_{ee} = 1.5\cdot 10^{-22}$ and 
$\epsilon_{\mu\mu} =\epsilon_{\tau\tau} = 1.1\cdot 10^{-22}$.

\section{Preliminary considerations \label{s-prel}}

In this section we introduce some approximations and discuss 
some features of 
cosmological evolution of neutrinos which can be helpful for determination
of numerical and analytical solutions of
kinetic equations and for description of the effects of neutrino
oscillations on BBN.

\subsection{Stationary point approximation \label{s-stat}}

Brute force numerical solution of integro-differential kinetic 
equations (\ref{dot-rss}-\ref{dot-rsa}) is rather difficult and
for large mass differences suffers from numerical instabilities,
especially in the resonance case. On the other hand, for small mass 
differences numerical calculations can be rather accurately performed.

In order to avoid numerical problems, in this work we have used an 
approximate but rather accurate method 
discussed in ref.~\cite{dolgov-itep,dolgov-nr}.
If damping terms, given by $\gamma_{\alpha\beta} \rho_{\alpha\beta}$,
are sufficiently large, differential equations for off-diagonal
components of density matrix can be formally solved in the stationary
point approximation. It simply means that r.h.s. of these equations
is assumed to be equal to zero and they reduce to linear algebraic
equations (see eqs. (\ref{rsa-1+1},\ref{sp_as}) below).
Such approximation works well when reaction rates are large,
much higher than expansion rate. The latter is true for sufficiently 
large mass differences of oscillating neutrinos. It can be 
shown~\cite{old-nus} (and it follows from the calculations presented 
below) that in non-resonance case, the maximum production rate of sterile 
neutrinos takes place at the temperature
\be
T^{\nus}_{prod} = (10.8,\,\,13.4)\, (3/y)^{1/3}\,
(\cos 2\theta)^{1/6}(\dm/{\rm eV}^2)^{1/6}\,\, {\rm MeV}
\label{tprodnus}
\ee
The first number above is for mixing of $\nus$ with $\nue$, 
while the second one is for mixing with $\num$ or $\nut$. Thus for
$\cos 2\theta \,  \dm > 10^{-6}$ eV${^2}$ 
the production of sterile neutrinos is efficient when $\Gamma \gg H$ 
and stationary point approximation is sufficiently accurate.

In this approximation one can express algebraically all off-diagonal
components of $(4\times 4)$ density matrix of oscillating neutrinos
through 4 diagonal components, $\raa$ ($a=e,\mu,\tau$) and $\rss$.
This is true if the off-diagonal components of effective potential
of active neutrinos can be neglected (see sec.~\ref{s-two+one}).
Found in this way expressions for off-diagonal components 
of the density matrix can be inserted into eqs.~(\ref{dot-rss}-\ref{dot-raa}) 
and we obtain a closed system of integro-differential equations 
for the diagonal components. The latter can be easily solved numerically 
as is done in the following sections. 

We remark that, while we have used quasi-stationary approximation for 
massive calculations, we have checked the validity of this approximation
by solving numerically the whole system of equations 
(i.e. without using the stationary point approximation) for several 
representative values of the parameters. This more exact approach takes much 
more computer time but we have not encountered any problem otherwise.
These more accurate results are always in a very good agreement
with those obtained with the approximate code.

\subsection{Kinetic equilibrium approximation \label{s-kin}}

Solution of kinetic equations is conceptually simpler in the case of
a large mass difference between neutrinos. 
In this case production of $\nus$ predominantly takes place 
at the temperatures 
(\ref{tprodnus}) which are much larger than the neutrino decoupling 
temperature (situation may be different for the resonance case, see below 
sec.~\ref{s-one+one}). Thus at least for non-resonance transition 
the distributions of active neutrinos should be close to the equilibrium one.
%\be
%\raa = f_{eq} =\left( e^{E/T} + 1\right)^{-1}
%\label{feq}
%\ee
%where we assumed that chemical potential is negligibly small (non-zero
%chemical potential is discussed in sec.~\ref{s-leptas}). 

We will estimate neutrino decoupling temperatures
in the Boltzmann approximation, i.e.
we will assume $f_{eq} = \exp(-E/T) = \exp(-y)$.
Under this assumption, if we take into account only
direct reaction terms in the collision integral, 
the evolution of the distribution
of non-oscillating neutrinos is governed by the equation:
\be
Hx\,{\partial f_\nu \over f_\nu\,\partial x} =
-{G_F^2 D\, y \over 3\pi^3 x^5 }
\label{fnue0}
\ee
where the coefficient $D$ is equal to $80(1+g_L^2 + g_R^2)$ if all
reactions in which neutrino can participate are taken into account. 
Here $g_L = \pm 1/2 + \sin^2 \theta_W$ and $g_R =
\sin^2 \theta_W$, plus or minus in $g_L$ stand respectively for $\nue$ or
$\nu_{\mu,\tau}$. The weak mixing angle $\theta_W$ is experimentally
determined as $\sin^2 \theta_W =0.23$.  Correspondingly
the decoupling temperatures with respect to the total reaction
rate are $T_{\nue} =1.34 $ MeV and $T_{\num,\nut} = 1.5 $ MeV. For this
and the following simplified estimates the thermally average value 
$\langle y \rangle = 3$ was taken.
%, but kinetic equations are solved 
%numerically without thermal averaging for arbitrary $y$. 

On the other hand, neutrino re-population, 
as one can see from eq.~(\ref{dot-raa}), is determined by the much weaker 
(inverse) annihilation rate. As a consequence,
in some range of temperatures, active neutrinos may
be in kinetic but not in chemical equilibrium. 
In Boltzmann approximation this corresponds to:
\be
\raa(x,y)=n_{a}(x)\exp(-y) \;\;\;\;\;\; a = e,\mu,\tau
\label{chem_eq}
\ee
In particular, if annihilation of active neutrinos into $e^+e^-$-pairs
is switched-off the total number density of active plus sterile
neutrinos is conserved. The rate of annihilation 
$\bar\nu \nu \lrar e^+e^-$ is given by eq. (\ref{fnue0})
with  $D= 16 (g_L^2+g_R^2)$ and the decoupling temperatures
would be $T^d_{\nue}=3.2$ MeV and $T^d_{\num,\nut} = 5.34 $ MeV.
Below these temperatures the total number density of all four neutrino 
flavors in comoving volume could not change. So extra neutrino species
are not created, though redistribution between $\nu_a$ and $\nus$ is
possible. 

In order to keep the computation time reasonably low, we assumed that, 
in the considered parameter range, the active neutrino distribution
functions are conveniently described by eq.~(\ref{chem_eq}).
For $\rss$ we did not made any simplifying assumption and took it
as an arbitrary function of time and energy. Let us remark
that, while we used kinetic equilibrium approximation for 
massive calculations, we 
have also solved numerically, for several values of the parameters,
the whole system of integro-differential
equations (i.e. assuming neither kinetic equilibrium
nor stationary point approximations). The results obtained are 
always in a very good agreement
with those obtained by using the approximate code.

\subsection{Effects on BBN \label{s-bbn}}

One should also keep in mind that BBN is especially sensitive to the
number density of electronic neutrinos because the latter directly
participate in reactions of neutron-proton transformation. A depletion
of $\nue$ and equally $\bar\nue$ number densities~\footnote{In the resonance
case $\nue$ and $\bar\nue$ can be depleted by a different amount and a 
noticeable charge asymmetry between active neutrinos can be created. 
This is discussed in sec.~\ref{s-leptas}}
through oscillations would give rise to a larger neutron freezing 
temperature and to a
larger abundances of primordial deuterium and helium-4, thus
enhancing the effect of the $\nu_s$ contribution into the energy 
density of the universe.
The described effect can be simply taken into account, if 
we assume that electron neutrinos are in kinetic equilibrium, that the
Boltzmann statistics is valid, and neglect the electron mass.
In this assumption, the neutron to proton interconversion rate 
$\Gamma_{np}$ is proportional to $\Gamma_{np}=(1+n_{e})/2$.
This means that the neutron freezing temperature scales as: 
\be
T_{f} \propto \left[\frac{g_{*}^{1/2}}{(1+ n_{e})/2}\right]^{1/3} 
\ee
where $g_{*}=10.75 + (7/4) \Delta N_
\nu$, and thus the global effect 
on BBN can be described as a variation of the effective number of
neutrino species:
\be
\Delta N_{\nu}^{\rm BBN} = 
\frac{4}{7} \left[\frac{10.75 + (7/4)\Delta N_{\nu}}
{((1+n_{e})/2)^{2}}-10.75 \right] 
\label{dnbbn}
\ee

If we consider possible deviations from kinetic equilibrium,
the same formulas apply but we have to replace $n_e$ by:
\be
\langle n_e \rangle = \frac{\int_0^\infty dy y^2 (y+\Delta m/T)^2 \, \ree(x,y)}
{\int_0^\infty dy y^2 (y+\Delta m/T)^2 e^{-y}}\,,
\label{ne-average}
\ee 
where $\Delta m =1.3 $ MeV is the neutron-proton mass difference and $T$ is the temperature.

If only one active and one sterile neutrinos were mixed, while mixing among
the active neutrinos was absent the described effect would lead
to a stronger restriction on the mixing angle in 
low $\dm$ region in the case of ($\nue$-$\nus$)-mixing.
In the case of 
$\num$ or $\nut$ mixing with $\nus$ the effect is less pronounced 
because depletion of $\nue$ is now a two step process. First, the
number density of $\num$ or $\nut$ decreases due to transition into
$\nus$ and after that some re-population of $\nu_{\mu,\tau}$ by
$\bar\nue \nue \rar \bar\nu_{\mu,\tau} \nu_{\mu,\tau}$ occurs
at the expense of the number density of $\nue$. This process however is
not very efficient because to be noticeable it should take place 
after decoupling of neutrinos from $e^+e^-$ plasma but practically 
simultaneously neutrinos decouple from themselves. However if the
active neutrino species are mixed then the depopulation
of $\nue$ would be much stronger as we see in what follows.
%On the other hand, if $\nus$ is predominantly mixed with $\nue$ 
%the account of the
%transformations between active neutrinos would lead 
%to a stronger  re-population
%of $\nue$ and the impact of $\nus$-$\nua$ mixing on BBN would be weaker.

\section{Mixed one active and one sterile neutrinos 
\label{s-one+one}}

As a first step let us consider the case when mutual mixing
of active neutrinos is absent and only one active neutrino is mixed with the 
sterile neutrino. Previous  BBN constraints on sterile neutrino 
admixture have been derived under this assumption.
We present below the results of our numerical calculations together
with some useful analytic estimates both for the non-resonance and
resonance cases.

\subsection{Non-resonance case \label{s-nores}}
The density matrix of the {\it oscillating} neutrinos is $2\times 2$.
It contains only four real functions: $\raa$, $\rss$, and 
$\ras=\ras^* = R + iI$, which have to be determined from the 
solution of kinetic equations:
\begin{eqnarray}
Hx \,\partial_x \raa
&=& 
i {\cal H}_{as}
\left(\rho_{as}-\rho_{sa}\right) - I_{coll}(q)
\label{kin2aa}
\\
Hx \,\partial_x \rss
&=&
-i {\cal H}_{as}
\left(\rho_{as}-\rho_{sa}\right) 
\label{kin2ss}
\\
Hx \,\partial_x \ras
&=& -i\left[\left({\cal H}_{aa}-{\cal H}_{ss}\right)-i \gamma_{as}\right]
\ras +i {\cal H}_{as}\left(  \raa - \rss \right)
\label{kin2as}
\end{eqnarray}
The formal solution of equation (\ref{kin2as}) is:
\be
\ras= i\int_0^x dx_1\frac{{\cal H}_{as} (\raa-\rss)_1}{(Hx)_1} 
\exp\left[ -i\int_{x_1}^x dx_2 \frac{\left({\cal H}_{aa} - {\cal H}_{ss} -i \gamma_{as}\right)_2}
{\left(Hx\right)_2}\right]
\label{rsa-1+1}
\ee
where the indices sub-1 and sub-2 indicate that the corresponding expressions 
are taken at $x_1$ or $x_2$. If $\gamma_{as}$ is large (more precisely we require 
$\gamma_{as}/H \gg 1$) then the integrals "sit" on the upper 
limit $x_1=x_2=x$ and in this way the stationary point result is obtained:
\be
\ras =  
\frac{{\cal H}_{as}}{({\cal H}_{aa}-{\cal H}_{ss})-i\gamma_{as}}
\left(\rho_{aa}-\rho_{ss}\right)   ~.
\label{sp_as}
\ee
This result is valid both for resonance and non-resonance cases if the mass 
difference is sufficiently large, but the limits of applicability are somewhat different.
As is mentioned above, for non-resonance case the temperature of $\nus$ 
production (\ref{tprodnus}) should be higher than the decoupling temperature
with respect to total reaction rate. This condition is satisfied for $\nue$ if 
$\dm > 3.7\cdot 10^{-6}$ eV$^2$ and for $\nu_{\mu,\tau}$ if
$\dm > 2\cdot 10^{-6}$ eV$^2$.
The resonance case is considered in the following section.

Substituting the result~(\ref{sp_as}) into eq.~(\ref{kin2ss}) we find:
\be
Hx \, \partial_x \rho_{ss} = 2
\frac{\gamma_{as}{\cal H}_{as}^2}
{({\cal H}_{aa}-{\cal H}_{ss})^2+\gamma_{as}^2} 
\, 
\left(\rho_{aa}-\rho_{ss}\right)
\label{sp_ss}
\ee
which, in terms of the mass difference, $\delta m^2$, and of the
vacuum mixing angle
\footnote{In our notations, 
$\dm$ is positive if sterile neutrino is heavier than 
active neutrino, in the limit $\theta\rightarrow 0$. Non resonance cases
correspond to positive $\delta m^2$.}
, $\theta$, can be rewritten as:
\be
Hx \, \partial_x \rho_{ss} = {\gamma_{a} \over 4}\, 
\left(\rho_{aa}-\rho_{ss}\right)\,
{ \sin^{2} 2\theta  \over 
\left( \cos 2\theta  - V_{aa} /\delta E \right)^2 +  
\gamma_{a}^2 /4\delta E^2 } 
\label{sp_ss_2}
\ee
From this expression (neglecting the $\gamma^2 $-term in the denominator)
follows that the rate of production of sterile neutrinos in
the primeval plasma is equal to:
\be
\gamma_s = {1\over 4} \,\gamma_a \tan^2 2\theta_m
\label{gammas}
\ee 
where $\theta_m$ is the effective mixing angle in matter
given by the last factor in eq.~(\ref{sp_ss_2}).
This result is twice smaller than simple estimates presented
in earlier papers~\cite{old-nus}-\cite{enqvist91}. 

In non-resonance case, when $\cos 2\theta \delta E  - V_{aa} \neq 0$  
(i.e. $\dm > 0$)
and one may neglect the $\gamma^2 $-term in the denominator, 
the number of additional neutrino species $\Delta N_{\nu}$ 
at BBN can be analytically calculated from 
eq.~(\ref{sp_ss_2}) in the limit of $\rss \ll \raa$ and $\raa=f_{eq}$. In this
approximation we obtain:
\be
(\dm_{\nue\nus}/{\rm eV}^2) \sin^4 2\theta^{\nue\nus} =
3.16\cdot 10^{-5} (\Delta N_\nu)^2
\label{dmess2}\\
(\dm_{\num\nus}/{\rm eV}^2) \sin^4 2\theta^{\num\nus} =
1.74\cdot 10^{-5} (\Delta N_\nu)^2
\label{dmmuss2}
\ee
If $\Delta N_{\nu}$ is not small, then 
$\rss$ in eq.~(\ref{sp_ss_2}) should
be included and a better approximation in the 
bounds above would be given by the factor $\ln^2 (1-\Delta N_{\nu})$
instead of $(\Delta N_{\nu})^2$. 

The described results are noticeably weaker than
earlier obtained analytical bounds 
They well agree, instead, with our numerical
calculations and with those of ref.~\cite{enqvist92}. 
Our numerical results are presented in figs.~\ref{fig1} 
($\num-\nus$ mixing) and \ref{fig2} ($\nue-\nus$ mixing). 
 In the first  panel, we show the effects of oscillations on 
the energy density of $\nu_e$. Specifically, we show 
the lines which correspond in the plane $(\sin^2 2\theta,\delta m^2)$
to fixed, but different, values of the energy density of electron 
neutrinos (normalized to the equilibrium value)
\footnote{Numerical calculations have been mostly done in the assumption 
of kinetic equilibrium for active neutrinos, as explained in sect.~\ref{s-bbn}.
In this assumption, the energy density of $\nu_{e}$ (normalized
to equilibrium value) is simply equal to the value of the overall normalization 
factor $n_{e}$ defined in eq.~(\ref{chem_eq}).}.
In the second panel, we show the energy density of sterile neutrinos.
In the third panel we show the total neutrino contribution 
to the energy density, each line corresponding to a certain value of
the number of extra neutrino flavors
$\Delta N_{\nu}$. If there was not an additional impact of $\nue$ on 
the neutron-proton 
transformations, these lines would present BBN bounds on mixing parameters. 
The total effect of oscillations on BBN, including the impact of $\nue$ on $n-p$ 
transformation according to eq.~(\ref{dnbbn}), is 
demonstrated in the fourth panel  
in terms of the effective number of neutrino species $\Delta N_{\nu}^{\rm BBN}$.
Finally, we show with red dotted lines the approximate analytical 
results described above. 
Numerical calculations were made by using stationary point
and kinetic equilibrium (for active neutrinos) approximation.
We checked by comparing with the complete calculations, that 
these approximations do not introduce significant errors.

As one can see, the bounds (\ref{dmess2},\ref{dmmuss2})
are reasonably accurate for large mass differences since,
in this case, re-population of active neutrinos is very efficient 
and we can always assume $\rho_{aa}= f_{eq}$.
On the contrary, when $\delta m^2$ is small (say,
$\dm\leq 10^{-3}$ eV$^2$) 
significant deviations are expected. In this case, sterile neutrino 
production takes place at low temperatures, see eq.~(\ref{tprodnus}), when 
inverse annihilation processes become slow with respect to 
the expansion rate of the universe. Slow re-population of active neutrinos
results in a decrease of $\Delta N_{\nu}$, since the production of $\nus$ goes 
at the expense of diminishing the number density of $\nua$. 

In the case of $\nue-\nus$ mixing, 
the decrease in the number density of $\nue$ and
$\bar\nue$ results in 
a decrease of the neutron-to-proton inter-conversion rates which
leads to a higher temperature of neutron freezing and to larger 
neutron-to-proton ratio and, as a result, to higher abundances  
of $D$ and $^4He$. 
This effect can be effectively described as an increase in $\Delta N_{\nu}^{\rm BBN}$
which is much larger than the simultaneous decrease of $\Delta N_{\nu}$ 
produced by the decrease of $\nue$ number density. This means that, for
small $\delta m^2$, the BBN bound 
on $\nue-\nus$ oscillations is considerably stronger than that 
obtained from the simple estimate given by eq.(\ref{dmess2}).
 
It is important to remark that in the case of $\num-\nus$ 
($\nut-\nus$) mixing, one expects a depletion of $\nue$ too. 
The number density of $\num$ (or $\nut$) goes down due to 
transformation of these particles into $\nus$. This opens window for 
$\bar \nue \nue \rar \bar \nu_{\mu,\tau} \nu_{\mu,\tau}$ 
and to a decrease of number density of $\nue$. 
In this case, however, the effect is considerably smaller and 
it has a magnitude comparable to the the decrease of $\Delta N_{\nu}$ 
produced by the decrease of $\num$ ($\nut$) number density.

%..........................................................
\subsection{Resonance case. Analytic estimates \label{s-res1}}
%......................................................

Situation in the resonance case (i.e. when $\dm < 0$)
is somewhat more complicated both from
the point of view of analytical and numerical calculations. 
We show, in this section, that simple analytical estimates 
can be obtained even in this case.

Let us return to kinetic equations~(\ref{kin2aa},\ref{kin2ss}). 
We assume that the collision integral $I_{coll}(q)$ 
has a negligible role during resonance. In other words, we assume that 
{\it during resonance} neutrinos are neither effectively created nor scattered
by other particles. This approximation is justified because the characteristic "time"
of collisions/annihilation, $\delta x_{coll} \sim \gamma/H$, is much longer 
than the inverse resonance width, $\delta x_{res} \sim \gamma/\delta E$, in 
the essential range of parameters.
We maintain, however, the decoherence term in the 
equations for the non-diagonal terms of the density matrix.
Under this assumptions, neutrino evolution {\it during resonance} 
is described  by the equations:
\begin{eqnarray}
Hx \, \partial_{x}(\rho_{aa}+\rho_{ss})
&=& 0 
\label{kin1-res}
\\
Hx \, 
\partial_{x}
(\rho_{aa}-\rho_{ss})
&=&
2i \, {\cal H}_{as}
\left(\rho_{as}-\rho_{sa}\right)
\label{kin2-res}
\end{eqnarray}
where $\ras = \rsa^{*}$ and the non diagonal 
term $\ras$ is given by eq.(\ref{rsa-1+1}).
We can write formally:
%.......................
\be
\ras(x,y)= 
i \exp(-F(x,y))
\int_0^x dx_1
\frac{\left[{\cal H}_{as} (\raa-\rss)\right]_1}
{(Hx)_1} 
\exp\left( F(x_1,y) \right)
\label{as-integral2}
\ee
%...........................................
where the function $F(x,y)$ is defined by the condition:
\be
Hx\, \partial_{x} F(x,y) 
= i \left({\cal H}_{aa} - {\cal H}_{ss} -i \gamma_{as}\right)
\label{f-def}
\ee
%.............
From the above expression, we see that the function $F(x,y)$ has 
stationary points $z_{\rm res}$ in the complex plane where:
\be
{\cal H}_{aa}(z_{\rm res},y) - {\cal H}_{ss}(z_{\rm res},y) 
-i \gamma_{as}(z_{\rm res},y) = 0
\ee 
We can explicitly solve this equation, obtaining:
\be
(z_{\rm res})_{n} = 
\left( \frac{y^2 \,(|C_v^{a}|+ i\epsilon_{as})}{|d_m|\cos(2\theta)}\right)^{1/6} 
\simeq 
\left( \frac{y^2 \,|C_v^{a}|}{|d_m|\cos(2\theta)} \right)^{1/6}
\left[ 1 + \frac{i}{6}\delta_{a} \right]
\exp \left(i \frac{\pi n}{3} \right) 
\ee
where $n = 0,1,...,5$, the parameters $d_m$, $C_v^{a}$ and 
$\epsilon_{as}$ are defined in
eqs.~(\ref{dE})-(\ref{gammaab}) and 
$\delta_{a}\equiv |\epsilon_{as}/C_v^{a}| \sim 10^{-2}$.
One can see that the usual resonance condition,
${\cal H}_{aa}(x_{\rm res},y) - {\cal H}_{ss}(x_{\rm res},y) = 0$,
gives the resonance point $x_{res}$:
\be
x_{\rm res} = 
\left( \frac{y^2 \,|C_v^{a}|}{|d_m|\cos(2\theta)} \right)^{1/6}
\label{res-time}
\ee
which is essentially equal to 
$x_{\rm res}\simeq{\rm Re}\left((z_{\rm res})_{0}\right)$.

We can take the integral~(\ref{as-integral2}) by using the saddle point approach:
%.........................
\begin{eqnarray}
\nonumber
\ras(x,y) &=& i \sqrt{2 \pi} \,
\exp\left[-F(x,y)+F(z_{\rm res},y)\right] \, \theta(x - x_{\rm res})\\
&\times&
\left[\frac{{\cal H}_{as}}{(Hx)}\right]_{\rm res} 
\left|\frac{\partial ^2 F}{\partial x^2}\right|_{\rm res}^{-1/2}
\exp\left(i\frac{\pi}{2}-i\frac{1}{2}\psi\right) \, (\raa-\rss)_{\rm res}
\label{as-saddle}
\end{eqnarray}
%....................
where the index $[{\rm res}]$ indicate that
the various quantities are evaluated at $z_{\rm res}=(z_{\rm res})_{0}$,
and the phase $\psi$ is given by 
$\psi = \arg(\partial^2 F/\partial x^2)_{\rm res}$.
We remark that the second derivative of the function $F(x,y)$ at resonance, 
can be expressed through the simple formula:
\be
\frac{\partial^{2} F(x,y)}{\partial x^{2}} = \frac{\left({\cal H}^{\rm vac}_{aa} - 
{\cal H}^{\rm vac}_{ss}\right)_{\rm res}}
{(Hx)_{\rm res}}
\frac{6i}{z_{\rm res}}
\ee
where ${\cal H}^{\rm vac}$ is the vacuum hamiltonian,
from which we see that $\psi\sim \pi/2$.

Equation~(\ref{as-saddle}) describes the approximate behavior of $\ras(x,y)$ as
a function of neutrino parameters and of the difference
$(\raa-\rss)_{\rm res}$ at the resonance.
We can use this expression to integrate eqs.~(\ref{kin1-res},\ref{kin2-res}). We have:
\begin{eqnarray}
(\raa+\rss) &=& {\rm const} 
\label{tot-saddle}
\\
\Delta(\raa-\rss) &=&  2i \int_{0}^{x} dx_0
\frac{({\cal H}_{as})_0}{(Hx)_0}
\left(\rho_{as}-\rho_{sa}\right)_0
\label{delta_saddle}
\end{eqnarray}
where $\Delta(\raa-\rss)$ is the total variation of 
$(\raa-\rss)$ across resonance. 
If we integrate the previous equations using, one more time, 
the saddle point approach, 
we obtain:
\be
\Delta(\raa-\rss) = - 2\pi 
(\raa-\rss)_{\rm res}
\left|\frac{\partial^2 F}{\partial x^2}\right|_{\rm res}^{-1}
\left[\left(\frac{({\cal H}_{as})_{\rm res}}{(Hx)_{\rm res}}\right)^2
\exp\left(i\frac{\pi}{2}-i\psi\right) + c.c.
\right]
\label{delta1-saddle}
\ee
This expression can be calculated explicitly, giving: 
\begin{eqnarray}
\nonumber
\Delta(\raa-\rss) &=& -  
\frac{\pi}{6}
\frac{|d_m| \sin^{2}(2\theta)}{ h \cos(2\theta)} \,
{\rm Re}\left(\frac{z_{res}^3}{y}\right) \,
(\raa-\rss)_{\rm res}
\\ 
&=& -
\frac{\pi}{6}
\frac{|d_m| \sin^{2}(2\theta)}{ h \cos(2\theta)}
\,
{\rm Re} 
\left( \frac{|C_v^{a}|+ i\epsilon_{es}}{|d_m|\cos(2\theta)}\right)^{1/2} \,
(\raa-\rss)_{\rm res}
\label{delta2-saddle}
\end{eqnarray}
As a final result,
we have thus obtained an algebraic relation between 
the value of the difference $(\raa-\rss)$
at the resonance, $(\raa-\rss)_{\rm res}$, and
its {\it total variation} 
$\Delta(\raa -\rss)$.
 
If we assume 
that $\Delta(\raa -\rss)$ is small, the value of 
$(\raa -\rss)$ at resonance is close to its initial value, i.e.
$(\raa-\rss)_{\rm res}\simeq (\raa)_{\rm in}$.
In this assumption, we immediately obtain from eqs.~(\ref{delta2-saddle},\ref{tot-saddle})
an estimate for active (sterile) neutrino depletion (production)
due to resonance. We have:
\begin{eqnarray}
P_{aa} &\equiv& \frac{(\raa)_{in}+\Delta \raa}{(\raa)_{in}} 
= 1-K
\label{our-aa}\\
P_{as} &\equiv& \frac{\Delta \rss}{(\raa)_{in}} = K 
\label{our-as} 
\end{eqnarray}
where
\be
K = \frac{\pi}{12}
\frac{|d_m| \sin^{2}(2\theta)}{ h \cos(2\theta)}
{\rm Re} 
\left( \frac{y^2 \,(|C_v^{a}|+ i\epsilon_{es})}{|d_m|\cos(2\theta)}\right)^{1/2}
\simeq 
A_{a} \left(\frac{|\delta m^{2}|}{ eV^{2}}\right)^{1/2} 
\frac{\sin^{2}(2\theta)}{\cos(2\theta)^{3/2}}
\ee
The coefficient $A_{a}$ is equal to $A_{e}=4.4 \cdot 10^{4}$
for $\nue-\nus$ mixing and $A_{\mu,\tau}=2.3\cdot 10^{4}$
for $\num-\nus$ or $\nut-\nus$ mixing.
The previous equations represent our final results.
A few comments are in order:
\begin{enumerate}
\item
%\par\noindent
These simple formulas, in the range of parameters in which
they are valid (i.e. small $K$ and, thus, small mixing angles), 
well agree with the results obtained 
by describing resonance effects by Landau-Zener approach \cite{LZ}.
We recall that, in this scheme, the active (sterile) neutrino 
survival (oscillation) probability $P_{aa}$ ($P_{as}$) is given by:
\begin{eqnarray}
P_{aa}&=& P_{\rm c}
\label{LZ-aa}\\
P_{as}&=& 1-P_{\rm c}
\label{LZ-as}
\end{eqnarray}
where:
\be
P_{\rm c} = \exp\left(-K \right)
\label{Pc}
\ee
and we assumed $\sin 2\theta \ll 1$.
The agreement between our simple formulas, which take 
into account decoherence effects, with the Landau Zener results 
(which do not take them into account), in the range of parameter 
in which our formulas are valid, makes us confident of the 
validity of eqs.~(\ref{LZ-aa},\ref{LZ-as}) 
in the whole parameter space. In the following 
we will use Landau-Zener formulas to obtain simple analytic 
predictions for the whole range of masses and mixing angles. 
\item
%\par\noindent
As also noted by \cite{enqvist92}, momentum dependence disappears from
eqs. (\ref{LZ-aa},\ref{LZ-as}) 
(and (\ref{our-aa},\ref{our-as})). 
This means that resonance does not produce, as a net result
\footnote{We remark that neutrinos with different momenta go through
resonance at different times,  see eq.~(\ref{res-time}). 
This means that, while we do not expect
strong distortions at the end of the process 
(i.e. after {\it all} relevant momenta have passed through resonance), 
a strong distortions can be generated at a given time as a result
of the fact that {\it only some} momenta, at that time, have reached resonance.},
a strong spectral distortion of neutrino  distributions.
As a consequence, the probabilities $P_{aa}$ and $P_{as}$
can be taken, in first approximation,
as global multiplicative factors of active and sterile neutrino
distribution functions.
\end{enumerate}

So far we have considered neutrino evolution through resonance.
Post resonance evolution is simple if the mixing angle is sufficiently 
small. It this case, after resonance, sterile
neutrinos are not produced. We have then:
\be
\frac{\rss(x,y)}{\raa(x,y)_{in}} = P_{as} = {\rm const}
\;\;\;\;\;\;\; x > x_{res} 
\label{ster-res}
\ee
On the other hand, active neutrino distribution function evolves due
annihilation and/or scattering, as described by the collision
integral $I_{coll}(q)$.
A simple description can be obtained if 
active neutrinos are approximately in kinetic equilibrium.
Under this assumption, the time evolution of the
normalization factor $n_{a}$ is described by the equation:
\be
\frac{dn_{a}(x)}{dx} = - \langle \gamma_{\rm ann} \rangle 
\left[ n_{a}(x)^{2}-1\right]
\ee
where the symbol $\langle  ...\rangle$ indicates averaging over
the Maxwell-Boltzmann distribution and  $\gamma_{\rm ann}$
is the total $\nu_{a} \overline{\nu}_{a}$ annihilation 
rate. If we assume that the deviation from chemical equilibrium
produced by resonance is generated 
abruptly at the time $\overline{x}_{\rm res}$ when neutrinos 
with momentum $y=3$ (which we take representative of the 
whole ensemble) goes through resonance, the previous equation 
can be solved explicitly giving the result:
\be
n_{a}(x) = \frac{1+A}{1-A}
\label{active-res}
\ee
where
\be
A = \frac{P_{aa}-1}{P_{aa}+1}
\exp\left[-\frac{2}{3}\langle \gamma_{\rm ann} \rangle
\left(\frac{1}{\overline{x}_{\rm res}^3}-
\frac{1}{x_{\rm BBN}^3} \right)\right]
\label{active-res-2}
\ee
and $x_{\rm BBN}^3$ is the "time" at which BBN takes place.

In figs.~\ref{fig3} and~\ref{fig4} we compare the analytical estimates 
with the numerical calculations. Specifically, in fig.~\ref{fig3} 
we compare, in the case of $\nue-\nus$ mixing,
the prediction of eqs.~(\ref{LZ-as},\ref{ster-res}) 
(red dotted lines)
with the energy density of sterile neutrinos obtained
from numerical calculations (solid lines)
as functions of the mixing angle $\sin^2 2\theta$,
for selected values of the mass difference $\dm$. 
One can see that eqs.~(\ref{LZ-as},\ref{ster-res})
correctly describe the dependence of sterile neutrino production on
the oscillation parameters over a wide range of masses and 
mixing angles. In particular, the analytical calculations are very
accurate at small mixings, while they tend to overestimate
sterile neutrino production at large angles.
We remark that numerical results shown in fig.~\ref{fig3} have been 
obtained with the complete code (i.e. without using stationary
point and kinetic equilibrium approximations)
in order to be sure not to introduce the errors of approximation into the calculation.
In fig.~\ref{fig4} we compare the analytical results (dotted lines) 
for $n_e$ with the energy density of $\nue$ obtained
from numerical calculations (solid lines). Blue dotted lines are obtained 
from eqs.~(\ref{LZ-as}) and do not take into account the post-resonance 
evolution of $\nue$. Red dotted lines include post resonance evolution 
according to eq. (\ref{active-res}). As one can see, analytic results 
give a correct qualitative understanding  of $n_{e}$ dependence
on oscillation parameters.

%............................................................
\subsection{Resonance case. Numerical calculations \label{s-res2}}
%............................................................

Stationary point approximation (\ref{sp_as}) is also valid
in the resonance case but
the region of validity is found in a different way because the 
effective production temperature is determined by the resonance 
condition and not by eq.~(\ref{tprodnus}).  

In order to discuss the validity of the stationary point 
approximation, we have to go back to eq.~(\ref{rsa-1+1}). 
The stationary point approximation follows 
from the expansion:
\begin{eqnarray}
\nonumber
i&\int_{x_1}^x& dx_2 
\frac{\left({\cal H}_{aa} - {\cal H}_{ss} -i \gamma_{as}\right)_2}
{\left(Hx\right)_2} \simeq\\ 
&\simeq& 
i \frac{\left({\cal H}_{aa} - {\cal H}_{ss} -i \gamma_{as}\right)_1}
{\left(Hx\right)_1} \;
(x-x_1) + \frac{i}{2} 
\frac{\partial \left[\left({\cal H}_{aa} - {\cal H}_{ss} -
i \gamma_{as}\right)/Hx\right]_1}
{\partial x}(x-x_1)^{2}
\label{expansion}
\end{eqnarray}
of the exponent in eq.~(\ref{rsa-1+1}). Due to the fact 
that $\gamma_{as}/H \gg 1$,
only a small region $\delta x = (x-x_1)/x_1 \simeq H/ \gamma_{as}$ 
is important. Thus the expansion can be truncated to first order
and the stationary point approximation can be easily obtained:
\be
\ras =  
\frac{{\cal H}_{as}}{({\cal H}_{aa}-{\cal H}_{ss})-i\gamma_{as}}
\left(\rho_{aa}-\rho_{ss}\right)
\label{sp2_as}
\ee
In resonance case the situation is more complicated. 
Far from the resonance, the above discussion applies without 
any modification. 
At the resonance point, instead, 
the factor $({\cal H}_{aa} - {\cal H}_{ss})$ vanishes, 
with the consequence that higher order terms in the expansion
may become important. This does not happen when:
\be
\left(\frac{x}{H}\right)_{\rm res}
\left[\frac{\partial ({\cal H}_{aa} - {\cal H}_{ss})}{\partial x}\right]_{\rm res} 
\frac{\delta x ^2}{2} 
\ll
\left(\frac{\gamma_{as}}{H}\right)_{\rm res} \delta x 
\ee
where $\delta x = (H/\gamma_{as})_{\rm res}$. In this case,
which corresponds $\dm > 10^{-1}\;{\rm eV}^2$ for $\nue-\nus$ mixing and 
$\dm > 10^{-3}\;{\rm eV}^2$ for $\num-\nus$ mixing,
the first order term in the expansion is anyhow dominant and 
eq.~(\ref{sp2_as}) continues to be valid.

For smaller masses, eq.~(\ref{sp2_as}) is not strictly true. 
Specifically, this equation overestimates $\ras$ at resonance. 
However, this does not introduce significant
errors for the evolution of the diagonal components, since at the
same time, resonance width is underestimated by a compensating
amount. 
The most direct way to understand this is to consider the equations
describing the evolution of the diagonal components 
which are obtained within the stationary point approximation:
\be
Hx \, \partial_x (\raa -\rss) = - 4
\frac{\gamma_{as}{\cal H}_{as}^2}
{({\cal H}_{aa}-{\cal H}_{ss})^2+\gamma_{as}^2} 
\, 
\left(\rho_{aa}-\rho_{ss}\right) - I_{coll}(\rho)
\label{dxrhoss2}
\ee 
Close to resonance, we can neglect the collision integral and
expand the r.h.s obtaining:
\be
\partial_x (\raa -\rss) =
- 4
\frac{(\gamma_{as}/Hx)_{\rm res}({\cal H}_{as}/Hx)^2_{\rm res} }
{B (x-x_{\rm res})^2 
+ (\gamma_{as}/Hx)_{\rm res}^2 } \left(\rho_{aa}-\rho_{ss}\right)
\label{sp-exp}
\ee
where 
\be
B \equiv 
\left[\frac{\partial ({\cal H}_{aa}-{\cal H}_{ss})/Hx}{\partial x}\right]_{\rm res}
\simeq -i \left[\frac{\partial ^{2}F(x,y)}{\partial x^{2}}\right]_{\rm res}~.
\ee
The index $[{\rm res}]$ indicates that the various quantities 
are evaluated at the resonance point and the function $F(x,y)$
is defined in the previous section by eq. (\ref{f-def}). 
If we define $\xi \equiv  B
(\gamma_{as}/Hx)^{-1} (x-x_{\rm res})$, 
eq.~(\ref{sp-exp}) becomes 
\be
\partial_\xi (\raa -\rss) =
- 4
\frac{({\cal H}_{as}/Hx)^2_{\rm res} }{B}
\frac{1}
{\xi^2 + 1 } \left(\rho_{aa}-\rho_{ss}\right)
\ee
One should note that the factor $\gamma_{as}$ has disappeared from
our formulas. We could expect then that eq.~(\ref{dxrhoss2}), 
derived under assumption of large $\gamma_{as}$,
is valid even in the limit of small $\gamma_{as}$. We can 
confirm this by integrating eq.~(\ref{dxrhoss2}) assuming a
slowly variation of  $\rho_{aa}-\rho_{ss}$. We obtain:
\be
\Delta (\raa -\rss) \simeq - 4 \pi
({\cal H}_{as}/Hx)^2_{\rm res} 
\left|\frac{\partial ^{2}F}{\partial x^{2}}\right|_{\rm res}^{-1}
\left(\rho_{aa}-\rho_{ss}\right)_{\rm res} 
\ee
This result coincides with the expected behavior, eq.~(\ref{delta1-saddle}), 
and confirm the validity of the stationary point approximation
in the resonance case even for small mass differences.

In fig.~\ref{fig5} ($\num-\nus$ mixing) and in fig.~\ref{fig6} 
($\nue-\nus$ mixing) we show our numerical results for the resonance case
in the plane $(\sin^2 2\theta,\dm)$. 
%These results have been obtained by using stationary point approximation. 
Let us remark that, besides the previous discussion,
we also have checked numerically the displayed results, by comparing them, 
for several values of the parameters, with the results of complete
numerical integration of kinetic equations. Similarly to fig.~\ref{fig1} and fig.~\ref{fig2} ,
the first panel shows the effect of oscillations on $\nu_e$ energy
density, the second panel shows the energy density of 
sterile neutrinos, the third panel describes the total neutrino contribution 
to the energy density and, finally, the fourth panel shows
the total effects of oscillations on BBN.

Our constraints on sterile neutrino admixture in the resonant case
are stronger than previous results of \cite{enqvist92},
which were obtained in the context of single momentum approximation
We are not able to understand the source of the observed difference.
We are however confident in the results of our numerical 
calculations, especially in the small mixing angles region,
where, as shown in the previous section, they can be
analytically understood with good accuracy.

In fig.~\ref{fig5} and \ref{fig6} 
we show with red dotted lines the prediction
obtained from eq.~(\ref{LZ-as}), which correctly reproduce 
(at small mixing) the numerical results. In Fig. \ref{fig6} we also 
show the prediction which can be obtained by using eqs.~({\ref{LZ-aa}})
and (\ref{active-res}) which give a qualitative understanding
of the behavior observed for $\Delta N_{\nu}^{\rm BBN}$.
 
%%%%%%%%%%%%%%%%%%%
%%%%%%%%%%%
\section{Mixed two active and one sterile neutrinos \label{s-two+one}}
%%%%%%%%%%%%%%%%%%

In the case that only two active neutrinos (say, $\nue$ and $\num$)
are mixed between themselves
and one sterile neutrino, all algebraic manipulations are relatively easy
and the resulting equations have a simple form and can be
explicitly written down below. We will consider large mass differences, so for 
the off-diagonal elements of the density matrix the stationary point approximation
may be used.

Using eq.~(\ref{dot-rab}) we can eliminate the $e\mu$-off-diagonal 
component of the neutrino density matrix:
\be
\rem = {\hem (\rmm - \ree )+ \hes \rsm - \hms \res \over
\hmm - \hee + i\gem }
\label{rem}
\ee
Substituting this expression into eqs.~(\ref{dot-rsa})
and their complex conjugate we obtain a closed system of equations
for four off-diagonal component $\rse$, $\rsm$, and their
complex conjugate $\res$ and $\rms$. They can be explicitly solved 
in terms of $\rmm$, $\ree$, and $\rss$. For example,
\be
\rse = \left( A C^* - B C \right)/ Det
\label{rse}
\ee
where $Det = |A|^2 - |B|^2$ is the determinant of the linear system of
4 algebraic equations for $\rsa$ and $\ras$ and
\be 
A &=& Z_{es}^* +{{\hms}^2 \over Z_{\mu e}} - 
{{\hem}^2 Z_{\mu e}^* \over Z_{\mu s}^* Z_{\mu e}^* - {\hes}^2} +
{ \left( \hms \hes\right)^2 \over Z_{\mu e}\, 
\left( Z_{\mu s}  Z_{\mu e} - {\hes}^2  \right)}
\label{A}\\
B &=& 2\, {\rm Im}\left[{\hms \hes \hem \over Z_{\mu s} Z_{\mu e}
- {\hes}^2 } \right] 
\label{B}\\
C &=& S_e - {\hem Z_{\mu e}^*  \over Z_{\mu s}^*Z_{\mu e}^*  - 
{\hes}^2 }\, S_\mu
+ { \hms \hes  \over Z_{\mu s} Z_{\mu e} - {\hes}^2 }\, S_\mu^*
\label{C} \\
Z_{\alpha\beta} &=& {\cal H}^{(1)}_{\alpha\alpha} -
{\cal H}^{(1)}_{\beta\beta} +i\gamma_{\alpha\beta}
\label{z-alphabeta}\\
S_e &=& {\hem \hms \over Z_{\mu e}}\, \left(\rmm - \ree \right)
+ \hes \left( \ree - \rss \right)
\label{Se}
\ee
$S_\mu$ and $\rsm$ can be obtained from eqs. (\ref{rem})-(\ref{Se})
by the interchange of $e$ and $\mu$.

Analytical expressions are particularly simple in the case of weak
mixing with $\nus$ and for sufficiently large mass differences, such
that refilling of active states is fast and their number densities 
approach equilibrium values, $\ree = \rmm = f_{eq}$. The 
analytical limits (\ref{dmess2}) and (\ref{dmmuss2}) are obtained
under similar assumptions. In this case the differential equation that
governs evolution of $\rss$ takes the form:
\be
\partial_x \rss =2\left( f_{eq} - \rss \right)\, {
 \ges \left( \hes \hmm - \hem \hms \right)^2 +
 \gms \left( \hms \hee - \hem \hes \right)^2 \over Hx\,
\left( \hee \hmm - {\hem}^2 \right)^2}
\label{dxrss2}
\ee
This result is valid in no-resonance case when $\hee \hmm -\hem^2 \neq0$.
If MSW-resonance exists then one should take into account imaginary parts 
proportional to coherence breaking terms $\gamma$. In numerical calculations
such terms have been included.

Simple expression (\ref{dxrss2}) has been derived under assumption of weak
mixing between active and sterile neutrinos. Relaxing this approximation
we find that resonance condition is:
\be
\dme \des \dsm - \hms^2\dms - \hes^2 \dse - \hem^2 \dem = 0
\label{res2+1} 
\ee
where $\Delta_{\alpha \beta} = {\cal H}_{\alpha \alpha} -
{\cal H}_{\beta \beta} $.

Eq. (\ref{dxrss2}) justifies neglecting of non-diagonal terms in the
effective potential. Indeed these non-diagonal terms are proportional to
the non-diagonal components of the active-active part of neutrino 
density matrix. As we see from eq. (\ref{dxrss2})
they should be compared to the diagonal ones. In the
situation close to thermal equilibrium $\rho_{ab} = 0$ if $a\neq b$, 
while $\rho_{aa} = f_{eq} \sim 1 $.

%%%%%%%%%%%%%%%%%%%%%%%%%%%%%%%%%%%%%%%%%%%%%%%%%%%%%%%%%%%%%%%%%%%%%%%%%%
\section{Realistic case of three active and one sterile 
neutrinos \label{s-three+one}}
%%%%%%%%%%%%%%%%%%%%%%%%%%%%%%%%%%%%%%%%%%%%%%%%%%%%%%%%%%%%%%%%%%%%%%%%%%%

Recent data from SNO \cite{sno}, KamLAND \cite{kamland},
and earlier data on solar \cite{solar-nu} and atmospheric \cite{atm-nu}
neutrinos have provided strong indications in favor of neutrino oscillations.
It is practically established now that all active neutrinos are mixed with
parameters given by the Large Mixing Angle (LMA) solution to solar neutrino
problem~\cite{holanda02}, see eq.~(\ref{lma})
%\begin{eqnarray}
%\dm_{12} &=& 7.3\cdot 10^{-5}\,\,\,{\rm eV}^2,\\
%\tan^2 \theta_{12} &=& 0.4
%\label{lma_angle}
%\end{eqnarray}
and by atmospheric neutrino data~\cite{atm-nu}, eq. (\ref{atm}).
%\begin{eqnarray}
%\dm_{23} &=& 2.5\cdot 10^{-3}\,\,\,{\rm eV}^2, \\
%\tan^2 \theta_{23} &\approx& 1
%\label{atmo_angle}
%\end{eqnarray}
Existence of fast transitions between $\nue$, $\num$, and $\nut$ 
may change BBN bound on mixing with sterile neutrinos especially
for small values of mass difference. This means that one has
to take into account mutual active neutrino mixing in the discussion of 
BBN bounds. Clearly the situation become much more 
complicated both conceptually and numerically. In the following, to
keep the discussion as simple as possible, we will make the
assumption that active-sterile neutrino mixing is small. 
 
In the case of 4 mixed neutrinos the algebra becomes significantly more complicated.
This is why we will not present explicit expressions for off-diagonal components of 
density matrix through the diagonal ones. All this work has been done by computer 
and the results have been inserted into differential equations for diagonal components.
The latter have been solved numerically. To check the accuracy of the procedure
we solved numerically the complete system of kinetic equations (\ref{dot-rss})-(\ref{dot-rsa})
for certain fixed values of the parameters
without any simplification and found that agreement with the approximate numerical results 
is very good. We did not do that for all interesting  values of the parameter because it
would take too much computer time, especially in the resonance case.

It is clear that, depending on masses and mixing, one can expect in 
general case one or more resonances among active and sterile neutrinos.
If $\nu_4$ is the heaviest mass eigenstate there may only be resonances in the
transformations among active neutrinos (they are quite efficient anyhow because of the
strong mixing). As $\nu_4$ is becoming lighter and lighter, there appear first one 
resonance in $\nua$-$\nus$ transformation, then two and ultimately three resonances.
The resonance condition is given by the vanishing of the determinant
of the $6\times 6$-matrix:
\be
D_6 = \det{
\left|
\begin{array}{llllll}
\,\,\,\,\,\dem & -\hmt  &\,\,\,\,\,\het &\,\,\,\,\,\hes & -\hms & 0   \\
-\hmt &\,\,\,\,\,\deta &\,\,\,\,\,\hem &\,\,\,\,\,0   & -\hts   &\hes \\
-\het &\,\,\,\,\,\hem  & \,\,\,\,\,\dmt & -\hts &\,\,\,\,\,0    &\hms \\
-\hes &\,\,\,\,\,0     & -\hts &\,\,\,\,\,\dms  &\,\,\,\,\,\hem &\hmt \\
-\hms & -\hts  &\,\,\,\,\,0  &\,\,\,\,\, \hem &\,\,\,\,\,\des   &\het \\
\,\,\,\,\,0  & -\hes  & -\hms &\,\,\,\,\,\hmt &\,\,\,\,\,\het   &\dts
\end{array}
\right|} = 0
\label{det31}
\ee
In the limit of weak mixing with $\nus$ the determinant of this matrix
is reduced to the product of two determinants:
\be 
D_6 \approx \det{ 
\left|
\begin{array}{lll}
\,\,\,\,\,\dem & -\hmt  &\,\,\,\,\,\het \\
-\hmt &\,\,\,\,\,\deta &\,\,\,\,\,\hem \\
-\het &\,\,\,\,\,\hem  & \,\,\,\,\,\dmt
\end{array}
\right|}\,\times
\det{
\left| 
\begin{array}{lll}
\,\,\,\,\,\dms  &\,\,\,\,\,\hem &\hmt \\
\,\,\,\,\, \hem &\,\,\,\,\,\des   &\het \\
\,\,\,\,\,\hmt &\,\,\,\,\,\het   &\dts
\end{array}
\right|}
\ee
The first term determines resonances in the active neutrino sector, while
the second one determines resonances between $\nus$ and three active 
neutrinos. Of course resonances manifest themselves in the process of
numerical calculations and we have checked that they are indeed in the right
positions determined by the above equations.

\subsection{Simplifying the problem}
If we assume that active-sterile neutrino mixing is small,
we can write the neutrino mixing matrix as:
\be
U =
\left(
\begin{array}{cccc}
U^{\rm act}_{e1} & U^{\rm act}_{e2} & U^{\rm act}_{e3} & \eta_1 \\
U^{\rm act}_{\mu 1} & U^{\rm act}_{\mu 2} & U^{\rm act}_{\mu 3} & \eta_2 \\
U^{\rm act}_{\tau 1} & U^{\rm act}_{\tau 2} & U^{\rm act}_{\tau 3} & \eta_3 \\
\epsilon_{1} & \epsilon_{2} & \epsilon_{3} & 1 
\end{array}
\right)
\ee
where $U^{\rm act}= U_{23}\cdot U_{13} \cdot  U_{12}$ is the $3\times3$ active neutrino 
mixing matrix and:
\begin{eqnarray}
\eta_{1} &=& -U^{\rm act}_{e1}\epsilon_{1} - U^{\rm act}_{e2}\epsilon_{2} - U^{\rm act}_{e3}\epsilon_{3}\\
\eta_{2} &=& -U^{\rm act}_{\mu 1}\epsilon_{1} - U^{\rm act}_{\mu3}\epsilon_{2} - U^{\rm act}_{\mu 3}\epsilon_{3}\\
\eta_{3} &=& -U^{\rm act}_{\tau 1}\epsilon_{1} - U^{\rm act}_{\tau 2}\epsilon_{2} - U^{\rm act}_{\tau 3}\epsilon_{3}
\end{eqnarray}
The previous relations are correct to first order in $\epsilon_1,\epsilon_2,\epsilon_3$
(or equivalently $\eta_1,\eta_2,\eta_3$).

We assume for simplicity that $\theta_{13} = 0$, while we take 
$\theta_{12} = \theta_{\rm sol}$ and $\theta_{23}=\theta_{\rm atmo}$. 
Concerning mass differences, we assume 
$\dm_{21}\equiv m_{2}^{2}-m_{1}^{2} = \dm_{\rm sol}$,
%and the values (\ref{atm}) 
%as reference values for 
%$\theta_{23}=\theta_{\rm atmo}$ 
and $\dm_{32}\equiv m_{3}^{2}-m_{2}^{2}=\dm_{\rm atmo}$.
Solar and atmospheric oscillation parameters are
given by eq.~(\ref{lma}) and~(\ref{atm}) respectively.

If $\theta_{13} = 0$, one can take advantage of the symmetry between $\num$ and $\nut$ 
to simplify the problem. The early universe, indeed, does not distinguish 
between $\nu_{\mu}$ and $\nut$ at $T<m_\mu$. We can thus
make a rotation in the flavor basis:
\be
\left(
\begin{array}{c}
\nue\\
\num'\\
\nut'\\
\nus
\end{array}
\right)
=
U_{23}^{-1}
\left(
\begin{array}{c}
\nue\\
\num\\
\nut\\
\nus 
\end{array}
\right)
\ee
In this basis the mixing matrix becomes:
\be
U' =
\left(
\begin{array}{cccc}
\cos\theta_{12} & \sin\theta_{12} & 0 & \eta'_1 \\
-\sin\theta_{12} & \cos\theta_{12} & 0 & \eta'_2 \\
0 & 0 & 1 & \eta'_3 \\
\epsilon_{1} & \epsilon_{2} & \epsilon_{3} & 1 
\end{array}
\right)
\ee
where:
\begin{eqnarray}
\eta'_{1} &=& \eta_{1} \;\;= \;\;-\cos\theta_{12} \cdot \epsilon_{1} - \sin\theta_{12} \cdot \epsilon_{2} 
\label{eta1}
\\
\eta'_{2} &=& \cos\theta_{23}\cdot\eta_{2}-\sin\theta_{23}\cdot\eta_{3} \;\;=\;\;  
\sin\theta_{12} \cdot \epsilon_{1} - \cos\theta_{12} \cdot \epsilon_{2} 
\label{eta2}
\\
\eta'_{3} &=& \sin\theta_{23}\cdot\eta_{2}+\cos\theta_{23}\cdot\eta_{3} \;\;=\;\; \epsilon_{3}
\label{eta3}
\end{eqnarray}
The problem which was written in terms of the parameters $(\eta_{1},\eta_{2},\eta_{3})$
can be easier treated in terms of the parameters $(\eta'_{1},\eta'_{2},\eta'_{3} = \epsilon_{3})$.
By using these parameters, it can be easily separated into two smaller sub-problems. 
In the following sections, we will discuss separately the constraint on $\eta'_{3}$
and $(\eta'_{1},\eta'_{2})$. In the final section we will combine our results.

\subsection{Bounds on $\eta'_{3}$}

The bounds on $\eta'_{3}=\epsilon_{3}$ can be immediately obtained from the
results of section~\ref{s-one+one}, since this is a 
pure ``1 active + 1 sterile neutrino'' problem.
Clearly the result depends on the value of the mass difference  
\be
\delta m^{2}_{43} = m_{4}^2 - m_{3}^2 ~.
\ee 
When $\delta m^{2}_{43} \ge 0$ active-sterile transitions are non-resonant. 
As a consequence, eq. (\ref{dmmuss2}) (and subsequent discussion) 
is valid. In terms of $\eta'_{3}$ we have thus:
\be
(\dm_{43}/{\rm eV}^2) \;(2\,\eta'_{3})^{4} =
1.74 \cdot 10^{-5} \ln^2(1-\Delta N_\nu)
\ee
When $\delta m^{2}_{43} < 0 $, sterile neutrino production
occurs though resonant transitions. 
This means that eq.~(\ref{LZ-as}) (and subsequent discussion) 
is approximatively valid. With very simple algebra, 
assuming that the number of extra neutrinos $\Delta N_{\nu}$
is approximatively given by the sterile neutrino energy 
density, we can obtain the following relation:
\be
(|\dm_{43}|/{\rm eV}^{2}) \, (2\,\eta'_3)^4
= 1.9 \cdot 10^{-9} \ln^2(1-\Delta N_\nu)
\ee
We remark that the two equations above 
take into account only sterile neutrino production, while
they do not consider the effects related
to depletion of other neutrino flavors. As a consequence,
 they are quite accurate for large 
mass differences, while they generally underestimate the effect on 
BBN at smaller masses (since they neglect effects of $\nue$ 
depletion on $n-p$ transformations). 
More precise bounds can be read 
from  fig.~\ref{fig1} and ~\ref{fig5} with the identification 
$\delta m^{2} \rightarrow \delta m^2_{43}$ 
and $\sin^{2}(2\theta) \rightarrow (2\eta'_3)^2$.

\subsection{Bounds on $\eta'_{1}$ and $\eta'_{2}$}

The problem described by the parameters
$(\eta'_{1},\eta'_{2})$ is much more complicated since it 
involves directly two active and one sterile neutrino, with mixing 
described by: 
\be
\left(
\begin{array}{c}
\nue\\
\num'\\
\nut'\\
\nus
\end{array}
\right)
=
\left(
\begin{array}{cccc}
\cos\theta_{12} & \sin\theta_{12} & 0 & \eta'_1 \\
-\sin\theta_{12} & \cos\theta_{12} & 0 & \eta'_2 \\
0 & 0 & 1 & 0 \\
\epsilon_{1} & \epsilon_{2} & 0 & 1 
\end{array}
\right)
\left(
\begin{array}{c}
\nu_{1}\\
\nu_{2}\\
\nu_{3}\\
\nu_{4} 
\end{array}
\right)
\label{2+1-mix}
\ee
The mixing angle $\theta_{12}$ and the mass difference
$\delta m^2_{21}$ are fixed by KamLAND and by solar neutrino 
experiments, see eqs. (\ref{lma}).

It is very difficult to obtain analytic estimates in the general case. 
However, one can simply understand what happen in the limiting cases.
Let us consider the situation $m_4 \gg m_2 > m_1$ 
(or $m_4 \ll m_1 < m_2$). In this case, we can approximatively
neglect the mass difference $\dm_{21}$. The angle $\theta_{12}$ can 
thus be rotated away, resulting in:
\be
\left(
\begin{array}{c}
\nue\\
\num'\\
\nut'\\
\nus
\end{array}
\right)
=
\left(
\begin{array}{cccc}
1 & 0 & 0 & \eta'_1 \\
0 & 1 & 0 & \eta'_2 \\
0 & 0 & 1 & 0 \\
\eta'_{1} & \eta'_{2} & 0 & 1 
\end{array}
\right)
\left(
\begin{array}{c}
\nu_{1}\\
\nu_{2}\\
\nu_{3}\\
\nu_{4} 
\end{array}
\right)
\ee 
By repeating the discussion of the previous section, one obtains then:
\begin{eqnarray}
(\dm_{41}/{\rm eV}^2) \;(2\,\eta'_{1})^{4} &=&
3.16 \cdot 10^{-5} \ln^2(1-\Delta N_\nu) 
\label{eta1-pos}\\
(\dm_{42}/{\rm eV}^2) \;(2\,\eta'_{2})^{4} &=&
1.74 \cdot 10^{-5} \ln^2(1-\Delta N_\nu)
\label{eta2-pos}
\end{eqnarray}
where we assumed positive mass differences, 
i.e. $\dm_{41} \simeq \dm_{42} \gg \dm_{21} > 0$. 
When mass differences are negative, i.e. 
$\dm_{41} \simeq \dm_{42} \ll \ -\dm_{21} < 0$, one expect 
instead
\begin{eqnarray}
(|\dm_{41}|/{\rm eV}^2) \;(2\,\eta'_{1})^{4} &=&
5.2 \cdot 10^{-10} \ln^2(1-\Delta N_\nu) 
\label{eta1-neg}\\
(|\dm_{42}|/{\rm eV}^2) \;(2\,\eta'_{2})^{4} &=&
1.9 \cdot 10^{-9} \ln^2(1-\Delta N_\nu)
\label{eta2-neg}
\end{eqnarray}

The other limiting situation that we can understand analytically
is $m_2 \gg m_4 \simeq m_1$. In
this case we can approximately neglect the mass difference 
$\dm_{41}$. The mixing in the $(1-4)$-sector can thus be 
re-absorbed, leading to:
\be
\left(
\begin{array}{c}
\nue\\
\num'\\
\nut'\\
\nus
\end{array}
\right)
=
\left(
\begin{array}{cccc}
\cos\theta_{12} & \sin\theta_{12} & 0 & 0 \\
-\sin\theta_{12} & \cos\theta_{12} & 0 & \varphi \\
0 & 0 & 1 & 0 \\
\varphi \cdot \sin\theta_{12} & -\varphi \cdot \cos\theta_{12} & 0 & 1 
\end{array}
\right)
\left(
\begin{array}{c}
\nu_{1}\\
\nu_{2}\\
\nu_{3}\\
\nu_{4} 
\end{array}
\right)
\ee 
where:
\be
\varphi = \eta'_1 \cdot \tan \theta_{12} + \eta'_{2} = 
-\epsilon_{2}/\cos(\theta_{12})
\label{phi-def}
\ee
We can see that the
mixing between active and sterile neutrinos is described by a
parameter $\varphi$. If $\varphi\neq 0$, a resonant transitions 
occur among $\nus$ and $\nu'_{\mu}$. This mechanism should dominate the production
of sterile neutrinos. As a consequence, we expect the following relation to 
be valid:
\be
(|\dm_{42}|/{\rm eV}^2) \;(2\,\varphi \cdot \cos(\theta_{12}))^{4} &=&
1.9 \cdot 10^{-9} \ln^2(1-\Delta N_\nu)
\label{phi-neg}
\ee
where $\varphi$ is given as a function of $\eta'_{1}$ and $\eta'_{2}$ by 
eq.~(\ref{phi-def})
\footnote{We note that the parameters $\eta'_{1}$ and $\eta'_{2}$ can be
chosen in principle in such a way that $\varphi = 0$. When $\varphi = 0$, 
sterile neutrinos are not mixed with active ones
and we should expect that sterile neutrinos are not produced. 
In reality, this result simply indicates that, in the limit considered
($m_{2}\gg m_{4}\simeq m_{1}$), no effects driven by the 
dominant mass difference, $\dm_{21}$
are expected. Even in this case, however, sterile neutrinos 
can in principle be produced, due to oscillation driven by sub-dominant 
mass differences $\dm_{41}$.}.

In order to check the validity of the previous conclusions and to
have a better understanding of the BBN bounds on sterile neutrinos, 
we solved numerically kinetic equations, varying, one at a time, the parameter $\eta'_{1}$
and $\eta'_{2}$. 
As a first step, we considered the situation 
when the mixing matrix $U$ can be written as $U = U_{24}\cdot U_{12}$, 
with $\dm_{21}$
and $\theta_{12}$ fixed at solar values, see eq.~(\ref{lma}). It is evident
that, in the limit of small mixing, one has $\theta_{24}\simeq \eta'_{2}$.
The results obtained are 
shown, as functions of $\dm_{42}$ and $\sin^2(\theta_{24})$, in
figs.~\ref{fig7} (for the case $\dm_{42}>0$) and in figs.~\ref{fig7bis} 
(for the case $\dm_{42}<0$). Red dotted lines correspond to analytic estimates
which are obtained from eq.~(\ref{eta2-pos}) in the case of positive $\dm_{42}$
 and from eq.~(\ref{eta2-neg}) in the case of negative $\dm_{42}$.
One can see that there is a good agreement with numerical calculations
for large values of mass difference. For small values of $|\dm_{42}|$
agreement is less satisfactory (especially for $\dm_{42}<0$) both because
the analytic estimates, derived for the limiting behaviours,
 are less appropriate
and because depopulation of electron neutrinos becomes relevant.
In particular, by comparing figs.~\ref{fig7} and~\ref{fig7bis} with
figs.~\ref{fig1} and~\ref{fig5} (which show the results obtained when
 we neglect the mixing $U_{12}$), one can see that generally the effect of
mixing between active neutrinos is to increase $\nue$ depletion,
 with the consequence of a stronger bound on the active-sterile mixing.

Similar conclusions can be obtained from figs.~\ref{fig8} and~\ref{fig8bis}.
In these figures we show numerical results for the situation in which 
the mixing matrix $U$ can be written as $U = U_{14}\cdot U_{12}$. 
In the limit of small mixing angle one has $\theta_{14}\simeq \eta'_{1}$.
Analytic estimates (red dotted lines) are obtained from 
eq.~(\ref{eta1-pos}) for $\dm_{41}\ge \dm_{21}$,
from eq.~(\ref{phi-neg}) for $0<\dm_{41} < \dm_{21}$ and 
from eq.~(\ref{eta1-neg}) for $\dm_{41} < 0$. 
One sees that there is a good agreement between numerical calculations and 
analytical estimates. In all the cases, analytic estimates 
underestimate the BBN bounds on active sterile mixing.

\subsection{Combining results}

As a summary, by considering variations of the parameter 
$\eta'_{3} = \epsilon_{3}$, we have obtained: 
\begin{eqnarray}
(\dm_{43}/{\rm eV}^2) \;(2\,\eta'_{3})^{4} &=&
1.74 \cdot 10^{-5} \ln^2(1-\Delta N_\nu)   
\;\;\;\;\;\;\; {\rm for} \;\;\;\;\;   m_{4} \ge m_{3} \\
(|\dm_{43}|/{\rm eV}^{2}) \, (2\,\eta'_3)^4 &=&
1.9 \cdot10^{-9} \ln^2(1-\Delta N_\nu)     
\;\;\;\;\;\;\;\; {\rm for} \;\;\;\;\;   m_{4}   < m_{3} 
\end{eqnarray}
These relations are approximate and generally underestimate the effect on 
BBN at small masses. 
More precise bounds can be read 
from fig.~\ref{fig5} with the identification 
$\delta m^{2} \rightarrow \delta m^2_{43}$ 
and $\sin^{2}(2\theta) \rightarrow (2\eta'_3)^2$.

By considering variations, one at a time, of $\eta'_{1}$ and $\eta'_{2}$
we have obtained:
\begin{eqnarray}
\nonumber
(\dm_{41}/{\rm eV}^2) \;(2\,\eta'_{1})^{4} &=&
3.16 \cdot 10^{-5} \ln^2(1-\Delta N_\nu) \\
\nonumber
(\dm_{42}/{\rm eV}^2) \;(2\,\eta'_{2})^{4} &=&
1.74 \cdot 10^{-5} \ln^2(1-\Delta N_\nu)  \\
&\;&\;\;\;\; {\rm for}    \;\;\;\;\; m_{4}  \ge m_{2} > m_1 \\
\nonumber\\
\nonumber
(|\dm_{42}|/{\rm eV}^2) \;
[2\,(\eta'_{1} \sin(\theta_{12}) +\eta'_{2}\cos\theta_{12})]^{4} &=&
1.9 \cdot 10^{-9} \ln^2(1-\Delta N_\nu) \\
&\;&\;\;\;\; {\rm for}    \;\;\;\;\; m_{2}  \ge m_{4} > m_1 \\
\nonumber\\
\nonumber
(|\dm_{41}|/{\rm eV}^2) \;(2\,\eta'_{1})^{4} &=&
5.2 \cdot 10^{-10} \ln^2(1-\Delta N_\nu) \\
\nonumber
(|\dm_{42}|/{\rm eV}^2) \;(2\,\eta'_{2})^{4} &=&    
1.9 \cdot 10^{-9} \ln^2(1-\Delta N_\nu) \\
&\;&\;\;\;\; {\rm for} \;\;\;\;\;   m_{2}  > m_{1} > m_4
\end{eqnarray}
As above, these constraints are approximate and 
more precise bounds can be read 
from figs. \ref{fig7},\ref{fig7bis},\ref{fig8},\ref{fig8bis}.

In general scenario all parameters are possibly different from zero
at the same time and the effects described above are all observed togheter. 
We do not expect, however, strong compensations to occur
 because the described effects have generally different magnitude 
and because, in terms of the parameter $(\eta'_{1},\eta'_{2},\eta'_{3})$, 
the problem is well separated in nearly all range of possible masses
(except for $ m_{2}  \ge m_{4} > m_1 $).
This means that the bounds 
on $(\eta'_{1},\eta'_{2},\eta'_{3})$ can be  translated 
into constraints on $(\eta_{1},\eta_{2},\eta_{3})$
and $(\epsilon_{1},\epsilon_{2},\epsilon_{3})$. We remind that
$\eta = U_{23} \cdot \eta'$ and $\epsilon = -(U_{12})^{-1} \cdot \eta'$.

\section{ Non-vanishing lepton asymmetry \label{s-leptas}}

We assumed above that lepton asymmetry of cosmological plasma is either
very small or has the natural value, $\eta \sim 10^{-9}$, of the same
order of magnitude as cosmological baryon asymmetry.
The charge asymmetric contribution to neutrino effective potential in 
plasma (eq.~(\ref{veff})) is
\be
V^{(asym)} = 1.1\cdot 10^{-20} (\eta^{(a)}/10^{-9})/ x^3
\label{vasym}
\ee
Thus one can see that for such $\eta$ the Hamiltonian (\ref{h-alphabeta}) 
is either dominated by the mass difference term (at low temperatures
$T= 1/x$) or by the non-local one (at high temperatures). However the 
magnitude of lepton asymmetry of neutrinos can be much larger than
$10^{-9}$ either due to resonance rise of the asymmetry because of charge
asymmetric transformation of $\nu_a$ into $\nus$ and $\bar\nu_a$ into 
$\bar\nus$~\cite{res-rise,dolgov-01-res} or due to primordially generated
large lepton asymmetry. There are plenty cosmological scenarios leading 
simultaneously to small baryon asymmetry and to much larger, even of order 
unity, lepton asymmetry~\cite{lept-asym}.
In both cases neglected up to this point charge asymmetric contribution 
into $V_{aa}$ (\ref{veff}) could become dominant.

The best bounds on the values of the lepton
asymmetry can be found from BBN, see review~\cite{dolgov02}. These bounds
have been considerably improved due to strong mixing between active 
neutrinos~\cite{dolgov-act,lunardini01}. It allows to put the upper
bound on the common chemical potential of all neutrino species 
\be
|\xi_a| =|\mu_a /T| < 0.07
\label{xia}
\ee 
This limit is loose enough to allow the charge asymmetric part of the 
potential to play a significant role in neutrino oscillations in the 
early universe. An analysis of BBN with non-zero 
chemical potentials and possible active-sterile neutrino mixing
has been performed recently in ref.~\cite{dibari03}.

Let us first consider naturally small initial value of 
$\eta\leq 10^{-9}$ and later more questionable larger values of primordial 
lepton asymmetry. For small asymmetry and positive mass difference between
$\nus$ and any active neutrino MSW-resonance would not be 
efficient (see below eq. (\ref{eta9}). Hence lepton asymmetry would remain always 
small and $V^{(asym)}$ (\ref{vasym}) may be neglected. Thus the bounds 
obtained above in non-resonance case are justified.
This may be not so if mass difference is negative and the resonance 
transition between $\nus$ and $\nu_a$ could take place. 
As a result a large lepton asymmetry, close to or even a little larger 
than $\eta =0.1$, can be generated in the active neutrino
sector~\cite{res-rise,dolgov-01-res}. Of course the total leptonic charge
of active plus sterile neutrinos is conserved and a lepton asymmetry 
of active neutrinos is compensated by opposite sign of lepton asymmetry 
of $\nus$. Neglecting $V^{(asym)}$ in this
case would be unreasonable. However, as we see below, a large lepton 
asymmetry can be generated only for very small values of mixing angle
between $\nus$ and $\nu_a$:
\be
\sin^4 2\theta \,|\dm| < \left\{ \begin{array}{ll} 
10^{-9}\,\,\,& {\rm for}\,\,\, \nue \\
3\times 10^{-9} \,\,\,& {\rm for}\,\,\, \nu_{\mu,\tau} 
 \end{array} \right.
\label{s4dm}
\ee 
Here mass difference is measured in eV$^2$. Note that the bound 
for $\nue$ is stronger than for $\nu_{\mu,\tau}$,
in contrast to  the bounds (\ref{dmess2}), (\ref{dmmuss2}).
For the values of parameters outside this region the approach used above
(with $V^{(asym)}$ neglected) is justified. As we have seen in the 
previous sections, the allowed interval of $(\theta-\dm)$ found  in the resonance
case but without  $V^{(asym)}$, is inside the parameter space restricted by (\ref{s4dm}). 
Hence we can conclude that the area
outside of (\ref{s4dm}) is definitely excluded and that the real bound
can be stronger. It is more difficult
to make conclusion about allowed range of the mixing parameters
inside the region (\ref{s4dm}) because an excessive number of neutrino
species may be compensated by electronic charge asymmetry. 
However, one should take into account that the generated lepton 
asymmetry could be strongly inhomogeneous~\cite{dibari-inhom} (see 
also~\cite{dolgov02,enqvist01}). If the mixing angle is larger 
than $10^{-4}$, so that the lepton asymmetry in the minimum (see below
discussion after eq.~(\ref{Sigma})) drops more than by 5 orders of magnitude,
then the universe would consist of chaotic domains with equal positive
and negative values of $\eta$~\cite{dolgov02} and the impact of the electronic
lepton asymmetry on BBN would be less pronounced after averaging over
such domains. On the other hand, as argued in ref.~\cite{enqvist01},
production of sterile neutrinos on
very steep domain walls might be resonance enhanced
and it may lead to unacceptable large $\Delta N_\nu$. The problem
is not yet settled down and deserves more investigation.

Anyhow, even the limit (\ref{s4dm}), which allows the
resonance generation of lepton asymmetry
is quite strong, much stronger than the non-resonant one and, for 
$|\dm|>10^{-4}$, it is also stronger than the 
limits obtained in direct measurements.

Resonance rise of lepton asymmetry does not take place 
in the case of strong ($\nus$-$\nu_a$)-mixing by the following evident 
reason. If mixing is large both active and sterile neutrinos quickly
reach thermal equilibrium when charge asymmetry is vanishing,
$\raa = \bar\raa = \rss = \bar\rss = f_{eq}$. To find the magnitude of
mixing angle when fast equilibration takes place 
we will use analytical calculations of ref.~\cite{dolgov-01-res}.
The evolution of the charge asymmetry 
is governed by the following equation derived in the limit of large 
values of parameters $K_a \gg 1$:
\be
&&{dZ_a(q)\over dq} = {10^{10} K^2_a (\sin 2 \theta)^2 \over 16\pi^2} 
\int_0^\infty dy f_{eq}(y) \nonumber \\
&&\int_{q_{in}}^q dq_1 \, \Sigma (q_1,y) \,
\exp \left[ - {\epsilon_a y \zeta \over qq_1} \right]
\sin\left[ \zeta \left({1\over y}-{y\over qq_1}\right)\right]
\sin\left[b_aK_a\int_{q_1}^q dq_2 {Z_a(q_2) \over q_2^{4/3}}\right]
\label{Z'}
\ee
where $\zeta=K_a(q-q_1) $ is effectively integration variable and
the constants depending upon the active neutrino flavor are given by:
\be 
K_e &=& 5.63\cdot 10^4 (|\dm|\,\cos 2\theta)^{1/2},\  \,\,\,\,\,
K_{\mu,\tau} = 2.97\cdot 10^4 (|\dm|\,\cos 2\theta)^{1/2}, \\
b_e &=& 3.3\cdot 10^{-3}(|\dm|\,\cos 2\theta)^{-1/3},\, \,\,\,
b_{\mu,\tau} = 7.8\cdot 10^{-3}(|\dm|\,\cos 2\theta)^{-1/3},\\
\epsilon_e &\approx& 7.4\cdot 10^{-3},\,\,\,\,\,\,\,\,\,\,\,\,\ \,
\ \ \ \ \ \ \ \ \ \ \ \ \ \ \ \ \ \ \,
\epsilon_{\mu,\tau} \approx 5.2\cdot 10^{-3}.
\label{consts}
\ee
The quantity $Z$ is related to charge asymmetry of active neutrinos
as
\be 
Z_a = 1.67 \cdot 10^9 \eta_a
\label{Za}
\ee
and the new ``time'' variable $q$ is chosen in such a way 
that in the limit of small charge asymmetry the resonance always
takes place at $q=y$, i.e. $q =\beta_a x^3$ with
\be
\beta_e = 6.63\cdot 10^3 \left(|\dm| \cos 2\theta\right)^{1/2}, \,\,\,
\beta_{\mu,\tau} = 1.257\cdot 10^4 \left(|\dm| \cos 2\theta\right)^{1/2}.
\label{xiemu}
\ee
The function $\Sigma (q,y)$ is given by
\be
\Sigma = \raa + \bar\raa - \rss -\bar\rss
\label{Sigma}
\ee

Equation (\ref{Z'}) was solved in ref.~\cite{dolgov-01-res} under 
assumption that $\Sigma$ varies very slowly and in an essential interval
of $q$ it is approximately equal to its initial value, $\Sigma_{in} =2$.
This approximation is valid for sufficiently small mixing angle $\theta$
and we will determine below when it is true. For small $q$ the charge
asymmetry $Z$ exponentially decreases and only for $q>1.278$ it starts
to rise. For $\sin 2\theta = 10^{-4}$ and $\dm =1$ eV$^2$ the asymmetry
drops by 3 orders of magnitude, for larger $\theta$ the decrease would
be much stronger.

In the limit of small charge asymmetry $\Sigma$ satisfies the equation
(in this equation and below we will omit sub-indices $a$):
\be
{d \Sigma \over dq} = - \left({K\sin^2 2\theta \over y}\right)^2\,
\int_{q_{in}}^q dq_1 \Sigma (q_1,y)\,
\exp \left[ - {\epsilon y \zeta \over qq_1} \right]\,
\cos \left[ \zeta \left({1\over y}-{y\over qq_1}\right)\right]
\label{Sigma'}
\ee
For large $K$ the integral in the r.h.s.``sits'' on the upper limit and 
can be easily taken. Thus we come to the equation:
\be
{d \Sigma \over dq} = - { K\epsilon\, y\,q^2 \sin^2 2\theta
\over \left( q^2 - y^2 \right) + \epsilon^2 y^4 } \,\,
\Sigma
\label{Sigma'2}
\ee
The solution of this equation is straightforward. It can be substituted
into eq. (\ref{Z'}) and we find that asymmetry does not rise if
$\Sigma$ quickly decreases. According to this solution, this happens if 
the mixing is bounded from below as given by eq. (\ref{s4dm}).

Let us now consider the case of a large {\it primordial} lepton asymmetry and
check if a stronger mixing between active and sterile neutrinos is allowed.
If so, then an experimental discovery of mixing between 
$\nu_a$ violating the bounds derived above would indicate
that the cosmological lepton asymmetry is much larger then the baryonic 
one or that neutrinos possess a new interaction, as e.g. coupling
to Majorons, which might strongly change the oscillation behavior in 
cosmological plasma~\cite{major}.

If the contribution of the lepton asymmetry into neutrino effective
potential (\ref{veff}) is non-negligible, then resonance transition
would exist for any sign of mass difference. In both cases, there are
two resonances but for $\dm >0$ these two resonances are either in
neutrino or antineutrino channels (depending on the sign of $\eta$), 
while for $\dm <0$ there is one resonance in neutrino channel and one
resonance in antineutrino channel. We assume that the mixing is 
sufficiently weak, so only the resonance transitions are of importance.
The resonance condition looks as following:
\be
5\cdot 10^{-13} c_2 \,\dm\, {x \over y} + 
10^{-20} \kappa_a\, {y\over x^5} 
\pm 1.1\cdot 10^{-20}\, \eta_9\, {1\over x^3} = 0
\label{res-asym}
\ee
where $c_2 =\cos 2\theta$, $\kappa_e = 1.137$, 
$\kappa_{\mu,\tau} = 0.317$, $\eta_9 = 10^9 \eta$, the signs $\pm$
refer to antineutrino ($+$) and neutrino ($-$) respectively.
Solving this equation we find the resonance values of neutrino momentum:
\be
y_{1,2} = \pm (0.55/\kappa) \eta_9\, x^2\,\left( 1 \pm \sqrt{1 - \lambda x^2}
\right)
\label{y12}
\ee
where $\lambda = 1.65 \cdot 10^8 c_2\, \kappa\, \dm /\eta_9^2$.

If $\dm > 0$, i.e. $\nus$ is heavier than active neutrinos,  
then the resonances occur only for $\lambda x^2 <1$.  
The oscillations could have an impact on BBN if they are efficient
in the interval $x=0.1 - 2$. Hence  the charge asymmetry should be
sufficiently large:
\be
\eta_9 > 1.3\cdot 10^4\, c_2^{1/2} \kappa^{1/2} (\dm)^{1/2}
\label{eta9}
\ee

We can roughly estimate resonance contribution into production of sterile
neutrinos integrating  eq.~(\ref{sp_ss_2}) over $y$ with 
$\rss \ll \raa = f_{eq}$:
\be
{d\over dx}\,\left({n_s \over n_{eq}}\right) = 
2.45 \cdot 10^{16} \kappa^{-1} s_2^2 c_2 (\dm)^2\, x^8\, 
{ y_1\,\exp (-y_1) + y_1\,\exp (-y_1) \over y_1 - y_2 }   
\label{ns/neq}
\ee
where $s_2 =\sin 2\theta$. 
Equation (\ref{sp_ss_2}) gives a good description of the
$\nus$ production rate also in the resonance case if the mass 
difference is not too small.
As we have mentioned above, both resonances 
are effective either in ($\nu_a$-$\nus$)-transition (for $\eta >0$)
or in ($\bar\nu_a$-$\bar\nus$)-transition (for $\eta <0$). Hence resonance
transitions diminish the absolute value of the original asymmetry. 
Indeed, if e.g. the resonances
operate in $\nua \lrar \nus$ channel the transition of $\nua$ into $\nus$
would be more efficient than the charge conjugated transition of
$\bar\nua$ into $\bar\nus$ and the number difference
between $\nua$ and $\bar \nua$ would decrease. 
If the resonance took place above $T=3-5$ MeV (or at $x$ below $x = 0.33 -0.2$),
then the distribution of $\nua$ and $\bar\nua$ would have equilibrium
form with chemical potentials satisfying $\xi_a = - \bar\xi_a$ and
the standard BBN code can be applied to calculations of light element
production. For smaller resonance temperatures chemical equilibrium
would not be restored and thus $\xi_a \neq - \bar\xi_a$. The code
should be modified to include this effect.

Let us find now what is the largest possible value of the lepton charge 
asymmetry $\eta$ allowed by the limit (\ref{xia}):
\be
\eta = {n_\nu - n_{\bar\nu} \over n_\gamma} =
{3\over 11}\,{\pi^2/6 \over 1.8}\,\xi = 0.25\, \xi
\label{eta-xi}
\ee 
where the first factor $3/11$ comes from the present day neutrino-to-photon 
number ratio and the second one comes from the ratio of the equilibrium 
number densities $(n_\nu - n_{\bar \nu})/(n_\nu + n_{\bar \nu})$. The
result is valid in the limit of small $\xi$. Since individual chemical potentials 
before equilibration enforced by $\nue-\num-\nut$
oscillations may be at most three times larger than the limiting value
$|\xi|=0.07$ we conclude that the maximum magnitude of the primordial lepton
asymmetry is $\eta = 0.05$. Another formally open possibility is large
values of $2\eta_{\nue} +\eta_{\num} + \eta_{\nut}$ and/or of
$2\eta_{\num} +\eta_{\nue} + \eta_{\nut}$ (and $\mu \lrar \tau$),
contributing into effective potential of different neutrino species
but a strong compensation in the sum $\xi_e+\xi_\mu+\xi_\tau$ down to 0.2.

If we take $\eta = 0.05$ then the contribution of the first resonance
is strongly suppressed and only the second resonance corresponding to 
smaller $y$ is effective. The produced $\nus$ would contribute to the
additional number of neutrino species as:
\be
\Delta N_\nu = 0.8\cdot 10^9 s^2_2 (\dm)^3 \int_0^{x_{max}} dx x^{10} 
\exp \left[-0.9\, \dm x^4\right]
\label{Delta-ns}
\ee
where it was assumed that $c_2 =1$. As an upper limit of integration we
will take $x_{max}^{(e)} =0.3$ 
in accordance with the estimates presented after eq.~(\ref{n-tot}).
If we mixing would be predominantly with $\num$ or $\nut$ one should
seemingly take $x_{max}^{(\mu,\tau)} =0.2$. However this is not so because 
refilling of active neutrino species would take place through
$e^+\, e^- \rar \nue\, \bar\nue$ with subsequent fast transformation
of $\nue$ into $\num$ and $\nut$ through oscillations and hence one should take
$x_{max}^{(\mu,\tau)} =x_{max}^{(e)}=0.3$.

%=====================

If the mass difference is smaller than $\dm_{max} \sim 10^2$ eV$^2$ 
(notice that cosmological limits on masses of active neutrinos~\cite{dolgov02} are 
not applicable to $\nus$ because their number density may be much smaller than the 
equilibrium one)
then $\exp [-0.9\dm x^4] \approx 1$ and we obtain
\be
\Delta N_\nu = 7\cdot 10^7 s^2_2\, (\dm)^3 x_{max}^{11} \approx 
10^2 s^2_2\, (\dm)^3
\label{Delta-N}
\ee
If $s_2 \sim 0.1$ then the oscillations could diminish primordial
lepton asymmetry of active neutrinos and produce noticeable amount of
$\nus$. 

However a more important for BBN effect comes from larger $x$. Above 
$x=0.3$ refilling of neutrino states may be neglected and the total 
number/energy
density of $\nue+\num+\nut+\nus$ would be constant. Correspondingly
production of sterile neutrino states would lead to an equal decrease of
active ones. Due to strong mixing between the later the deficit of 
active neutrinos would be equally distributed between all three active
species. As we have discussed above, BBN is very sensitive to a deficit 
of $\nue$. We estimate the $\nus$ production using eq.~(\ref{Delta-ns})
with $x_{max} = 1.4$, i.e. inverse temperature of neutron-proton freezing. 
For $\dm =1$ eV$^2$ we obtain $\Delta N_{\nus} = 3\times 10^8 s_2^2$
and for small $\dm$, such that $\dm\,x^4 < 1$,
$\Delta N_{\nus} = 3\times 10^9 s_2^2 (\dm)^3$. 
The deficit of $\nue$ (or $\bar\nue$, depending on the sign of initial
$\eta$) would be equal to one third of that. However there is one more 
twist, namely there was some primordial lepton asymmetry of active
neutrinos and oscillations between $\nua$ and $\nus$ would diminish the
absolute magnitude of the latter. As a result we could arrive to (almost)
equal number densities of active neutrinos and antineutrinos which would
be suppressed with respect to the standard one by the factor
$\exp [-|\sum_a \xi_a^{(in)}|/3]$ where $\xi_a^{(in))}$ are primordial
values of chemical potentials of active neutrinos. According to
eq.~(\ref{dnbbn}),  a deficit of
$\nue$ and equal deficit of $\bar\nue$ is equivalent to addition of
$\Delta N_\nu = 12.3 |\xi|$. In this way we obtain essentially the 
same bound as eq. (\ref{xia}). However, it is not so in a general
case. Invoking an additional parameter, the value of primordial lepton
asymmetry, evidently makes the bounds on mixing between $\nus$ and
$\nua$ less restrictive as well as the bounds on the value of 
lepton asymmetry obtained from BBN in absence of mixing with $\nus$.

Now let us consider negative mass difference, $\dm <0$,
when resonance is possible even with vanishingly small charge asymmetry.
Since active neutrinos are assumed to be heavier than $\nus$ the
mass difference is bounded by the recent WMAP 
data~\cite{wmap} by $\dm <0.06$ eV$^2$. In this case, of negative 
$\dm$,  there are again two resonances but now one is in neutrino
channel, while the other is in antineutrino channel, as can be seen
from eq.~(\ref{y12}) with $\lambda <0$. If the initial value of $|\eta|$ is sufficiently large, 
then $|\eta|$ would rise 
as a result of oscillations between $\nua$ and $\nus$ in contrast to 
positive mass difference, $\dm >0$, when oscillations lead to a decrease
of $|\eta|$. 

The rate of transformation of $\nua$ into $\nus$ and of the charged 
conjugated process can be found from equation which is similar to 
eq.~(\ref{ns/neq}) where only one resonance is effective. For a positive
lepton asymmetry, $\eta >0$, the resonance in neutrino channel takes 
place at $y = y_1$ (with positive sign of the square root) and the 
resonance in antineutrino channel takes place at $y=y_2$. 
For a large and, say, positive $\eta$ the resonance in neutrino
channel is too far, $y_1\gg 1$, in the
essential range of $x$ and its contribution can be neglected - this is
similar to the discussed above case of positive $\dm$. The resonance
in antineutrino channel takes place at 
$y_2 \sim (4.5\cdot 10^7/\eta_9) x^4$ and could efficiently transform
$\bar\nua$ into $\bar\nus$. It would lead to an increase of $\eta$.
For a more accurate calculation of the transformation rate one should
include non-vanishing chemical potential into number density of active
neutrinos in eq.~(\ref{sp_ss_2}), $\raa =f_{eq}= \exp (-y+\xi)$ but 
since we usually consider small $\xi \ll 1$, this correction is not
of much importance. The oscillations  create an increase of
the absolute value of the lepton asymmetry and hence the BBN bounds 
on the initial magnitude of the latter should be generally stronger than in 
no-oscillation case. However there are several effects ``working'' in the 
opposite direction
which may compensate each other and so the bounds on $\eta$ for concrete 
values of the oscillation parameters in each case should be derived 
separately. It can be approximately done using results presented in
this section.

\section{Conclusion\label{s-concl}}

\begin{figure}[hbt]
\begin{center}
\epsfig{file=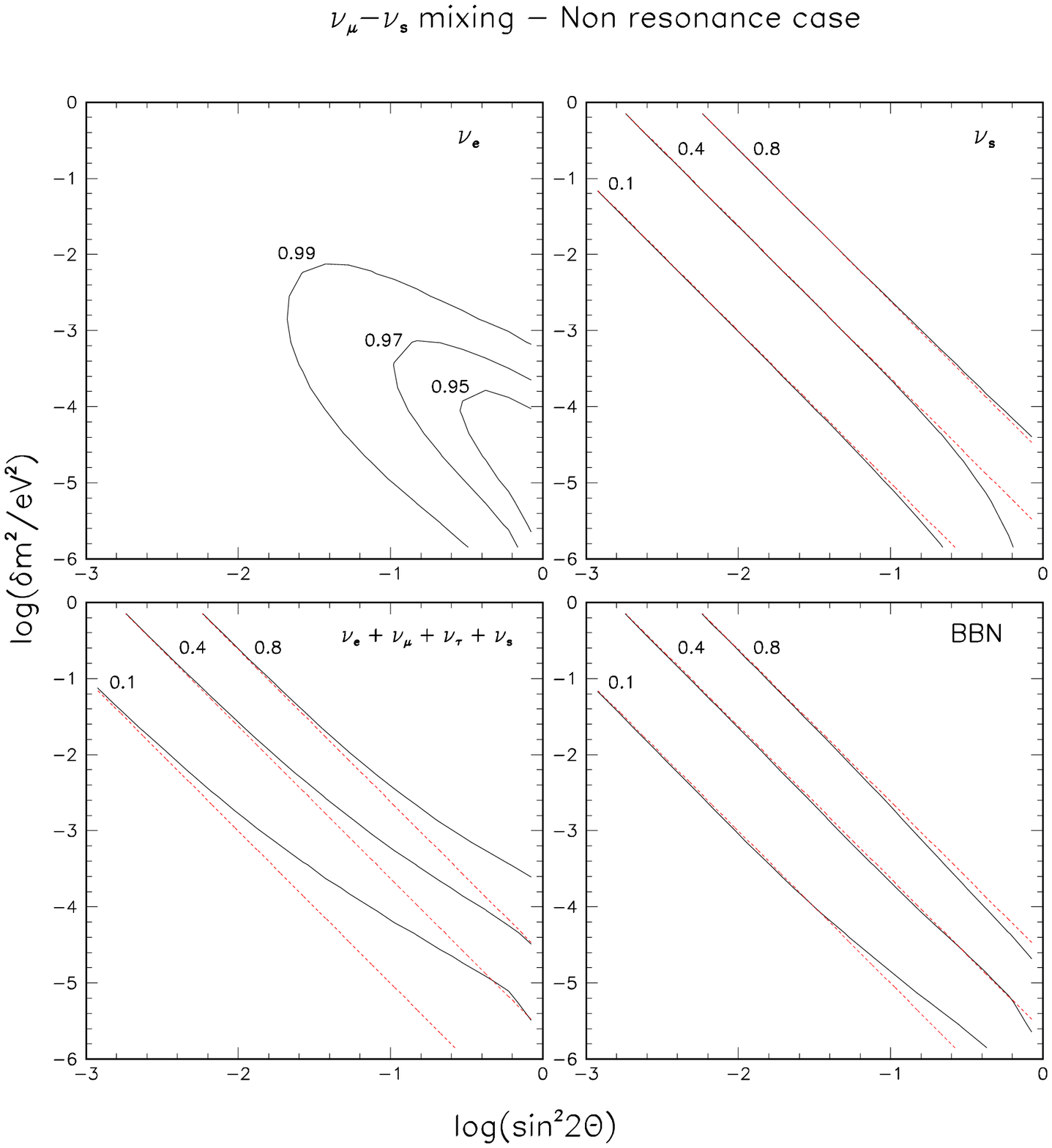,height=18cm,width=15cm}
\vspace{-1.2cm}
\end{center}
\bigskip
\caption{Numerical results for $\num-\nus$ mixing. Non resonance case.
{\it First panel:} Iso-countour lines for $\nue$ energy density. Each line
corresponds to fixed value of $\nue$ energy density.
The indicated values are normalized to equilibrium value. 
{\it Second panel:} Iso-contour lines for $\nus$ energy density.
{\it Third panel:} Iso-contour lines for total neutrino contribution 
to the energy density. Each line corresponds to a fixed value of the
number of extra neutrino flavors $\Delta N_{\nu}$.
{\it Fourth panel:} Total effect of neutrino oscillation on BBN. Each line
corresponds to a fixed value of the effective number of neutrinos
$\Delta N_{\nu}^{\rm BBN}$ species calculated according to 
eq.~(\ref{dnbbn}). Experimental bounds on $\Delta N_{\nu}^{\rm BBN}$ 
exclude the region above the corresponding curve.
Red dotted lines correspond to analytical estimates obtained from
eq.~(\ref{dmmuss2}) and following discussion.
The quantities presented in this and in the following figures are
taken at the BBN ``time'', $x_{\rm BBN}=1.4$.
}

\label{fig1}

\end{figure}

\begin{figure}[hbt]
\begin{center}
\epsfig{file=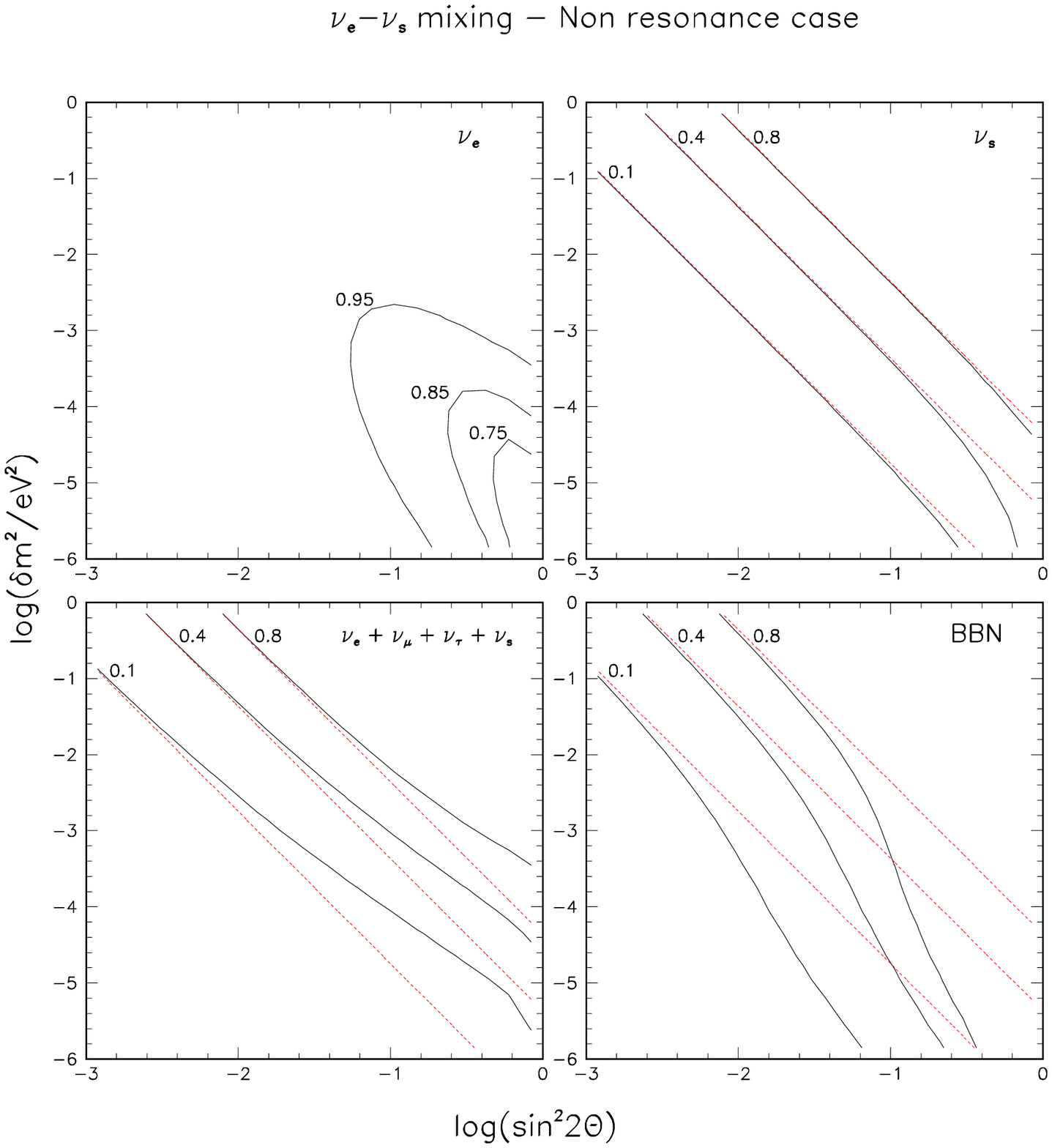,height=18cm,width=15cm}
\end{center}
\bigskip
\caption{Numerical results for $\nue-\nus$ mixing, non resonance case.
See Fig.1 for detailed explanation of the various panels and 
for a definition of the various lines.
Red dotted lines correspond to 
analytical estimates obtained from eq.(\ref{dmess2}) and following 
discussion.}
\label{fig2}
\end{figure}

\begin{figure}[hbt]
\begin{center}
\epsfig{file=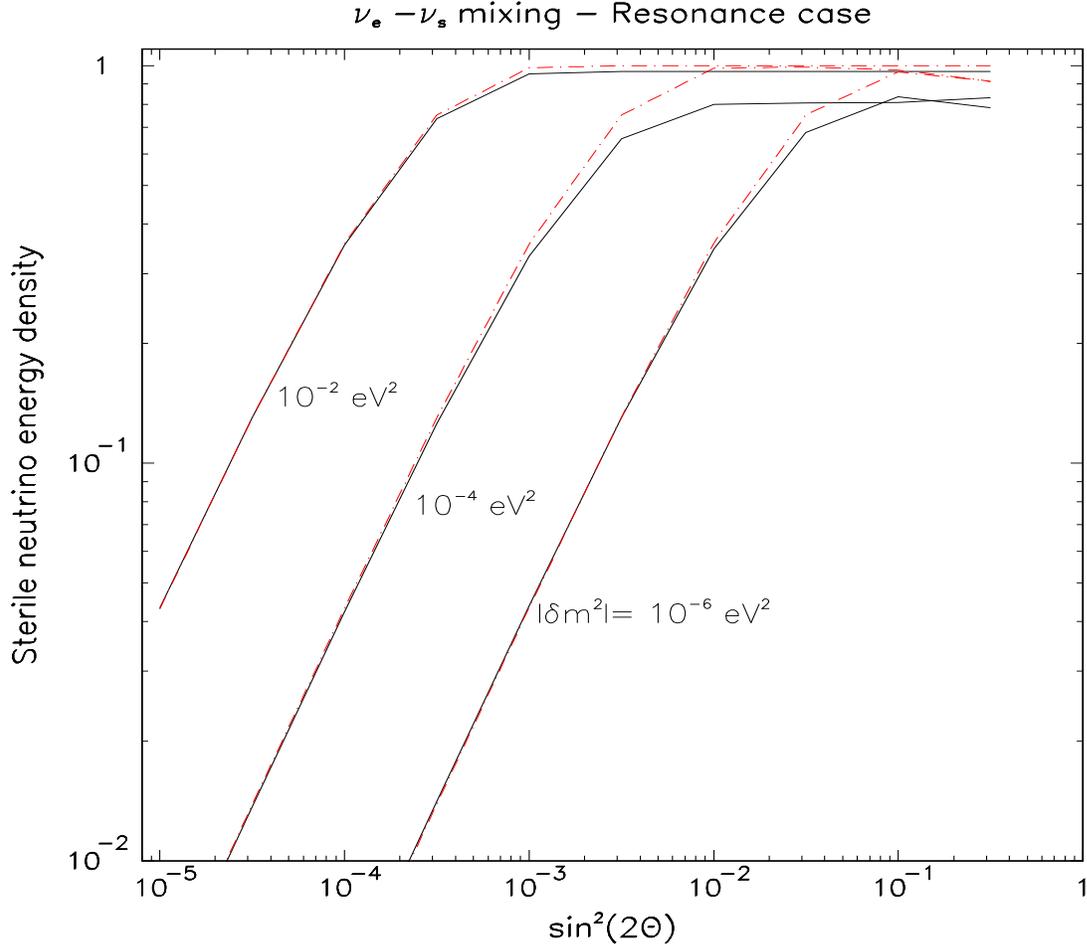,height=18cm,width=15cm}
\end{center}
\bigskip
\caption{Comparison between numerical results (solid lines) and analytic
estimates (red dotted lines) for $\nue-\nus$ mixing, resonance case.
The various lines show the energy density of sterile neutrinos (at the time
of BBN) as a function of the mixing angle $\sin^{2}(2\theta)$, 
for selected values of the mass difference $\dm$. Analytic estimates are 
obtained from eq.~(\ref{LZ-as}) and following discussion.}
\label{fig3}
\end{figure}

\begin{figure}[hbt]
\begin{center}
\epsfig{file=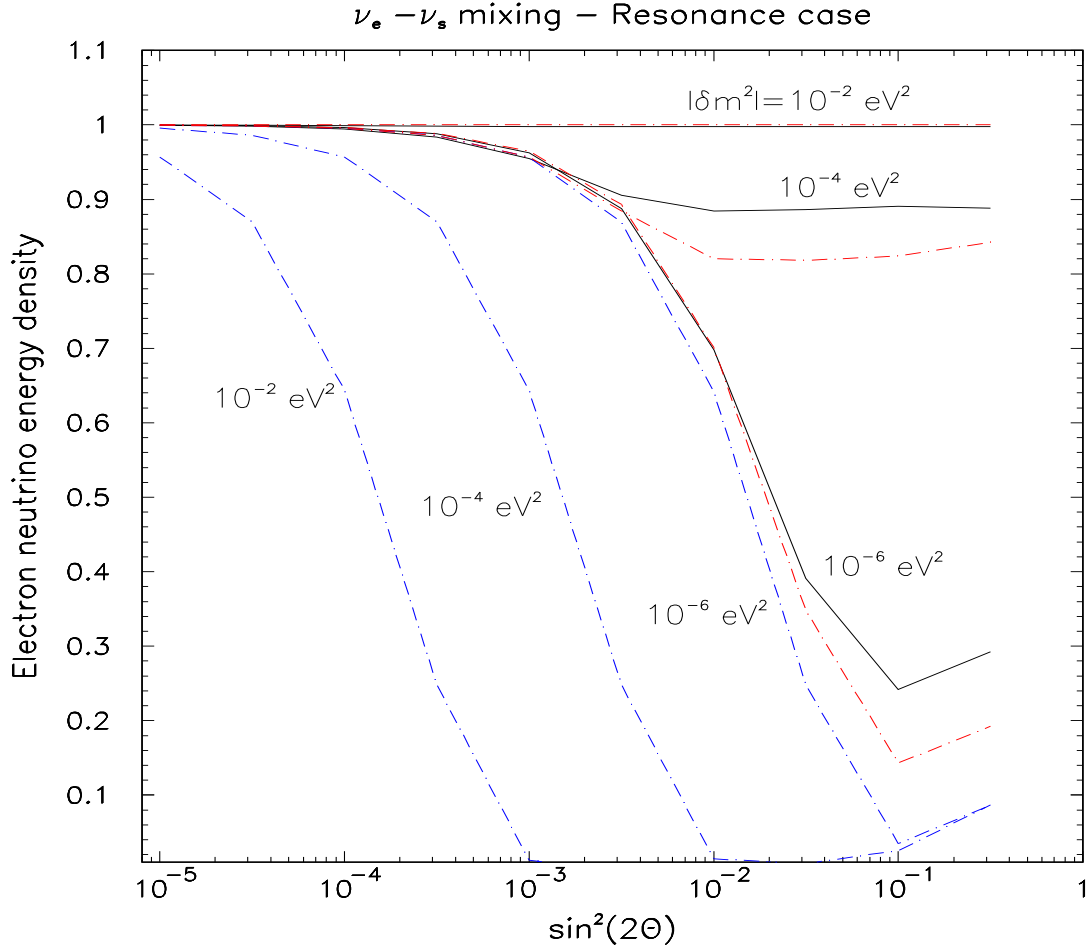,height=18cm,width=15cm}
\end{center}
\bigskip
\caption{Comparison between numerical results (solid lines) and analytic
estimates (dotted lines) for $\nue-\nus$ mixing, resonance case.
The various lines show the energy density of electron neutrinos (at the time
of BBN) as a function of the mixing angle $\sin^{2}(2\theta)$, 
for selected values of the mass difference $\dm$. Blue dotted lines,
which are obtained from eqs.~(\ref{LZ-aa}), 
do not take into account $\nue$ post-resonance evolution. Red dotted
lines take it into account following eqs.(\ref{active-res},
\ref{active-res-2}).}
\label{fig4}

\end{figure}

\begin{figure}[hbt]
\begin{center}
\epsfig{file=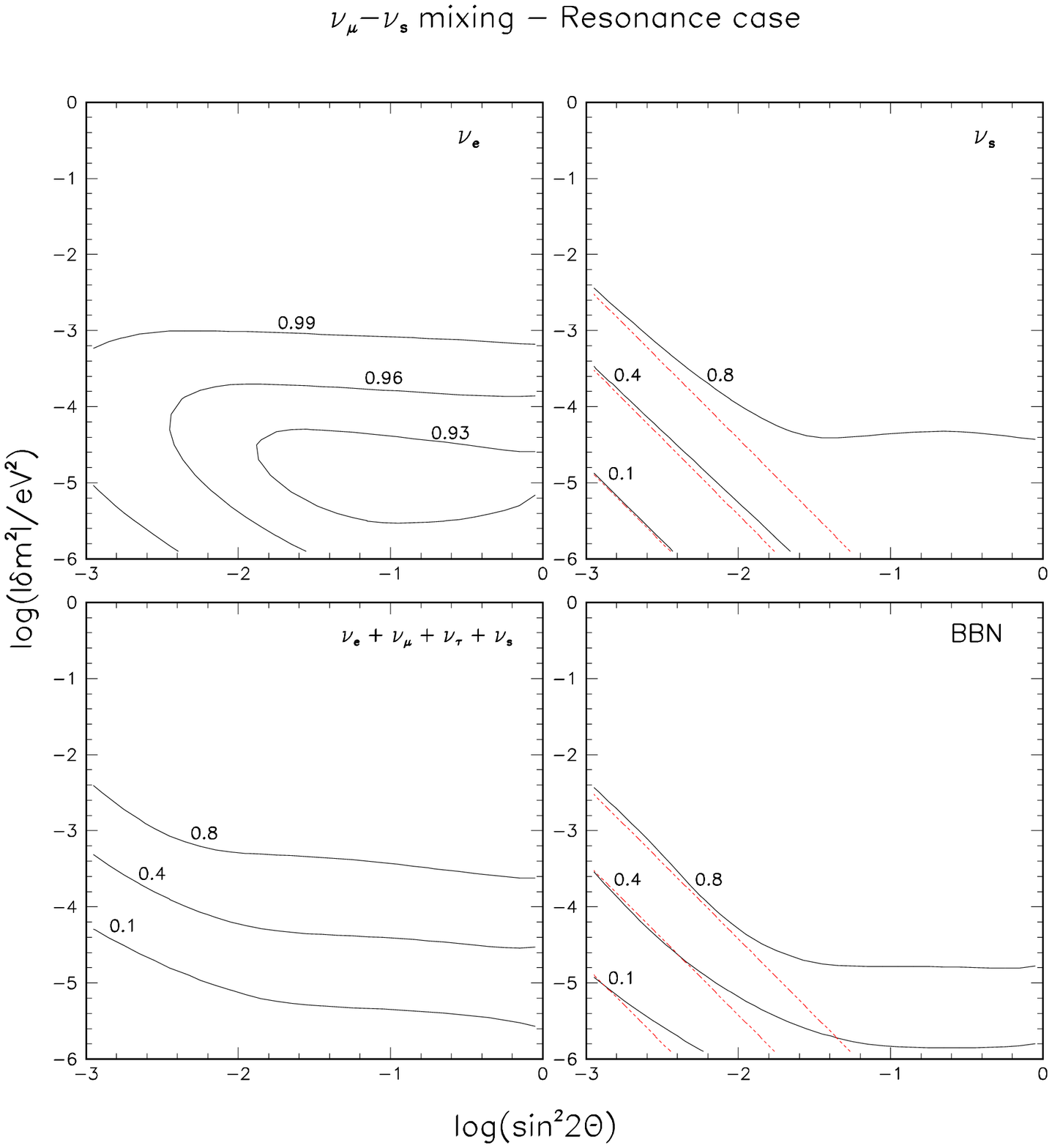,height=18cm,width=15cm}
\end{center}
\bigskip
\caption{Numerical results for $\num-\nus$ mixing, resonance case.
See Fig.1 for detailed explanation of the various panels and 
for a definition of the various lines.
Red dotted lines correspond to analytical estimates obtained 
from eq.(\ref{LZ-as}) and following discussions.}
\label{fig5}
\end{figure}

\begin{figure}[hbt]
\begin{center}
\epsfig{file=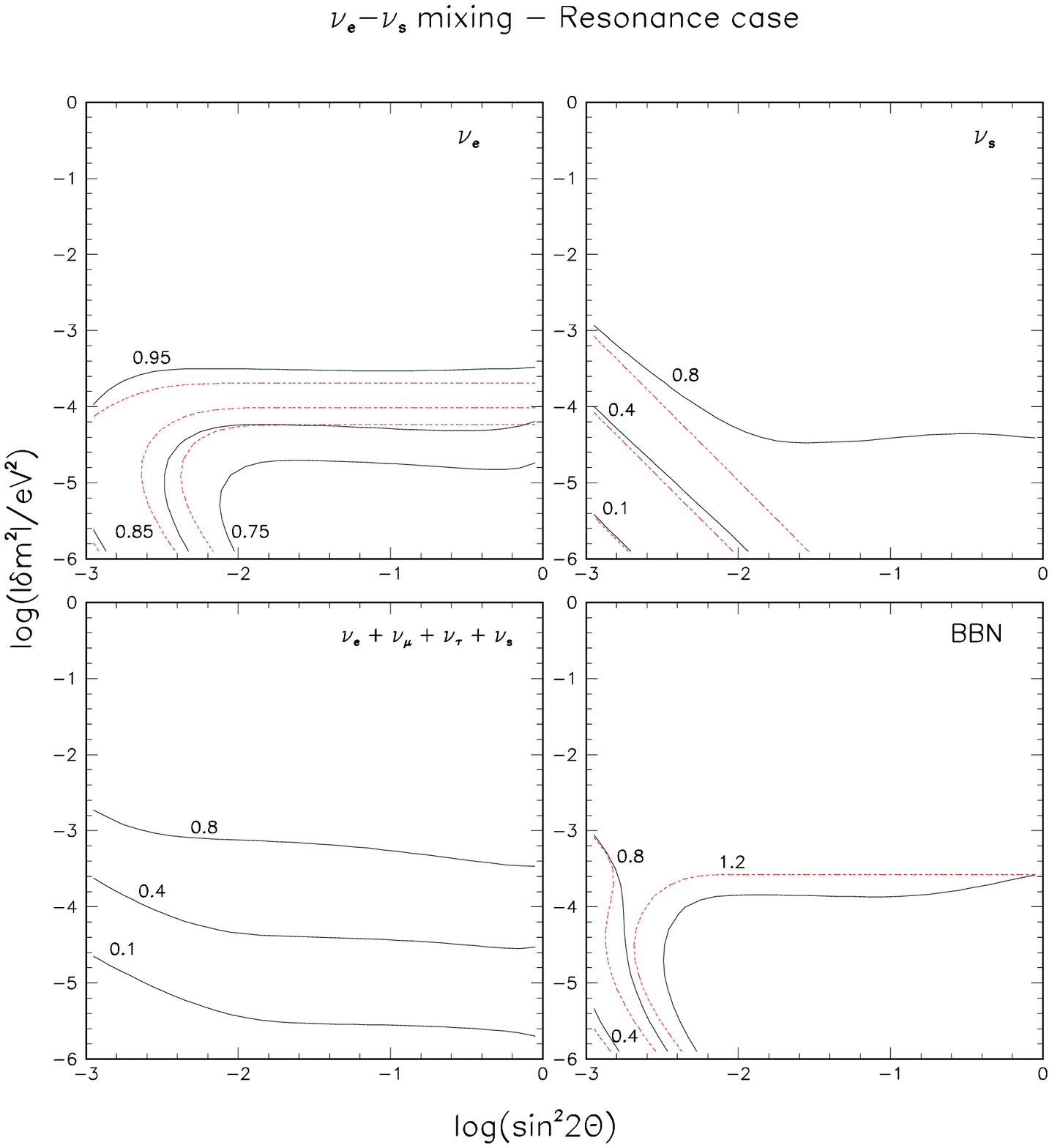,height=18cm,width=15cm}
\end{center}
\bigskip
\caption{Numerical results for $\nue-\nus$ mixing, resonance case.
See Fig.1 for detailed explanation of the various panels and 
for a definition of the various lines.
Red dotted lines correspond to analytical estimates 
obtained from eq.(\ref{LZ-as}) and following 
discussions.}
\label{fig6}
\end{figure}

\begin{figure}[hbt]
\begin{center}
\epsfig{file=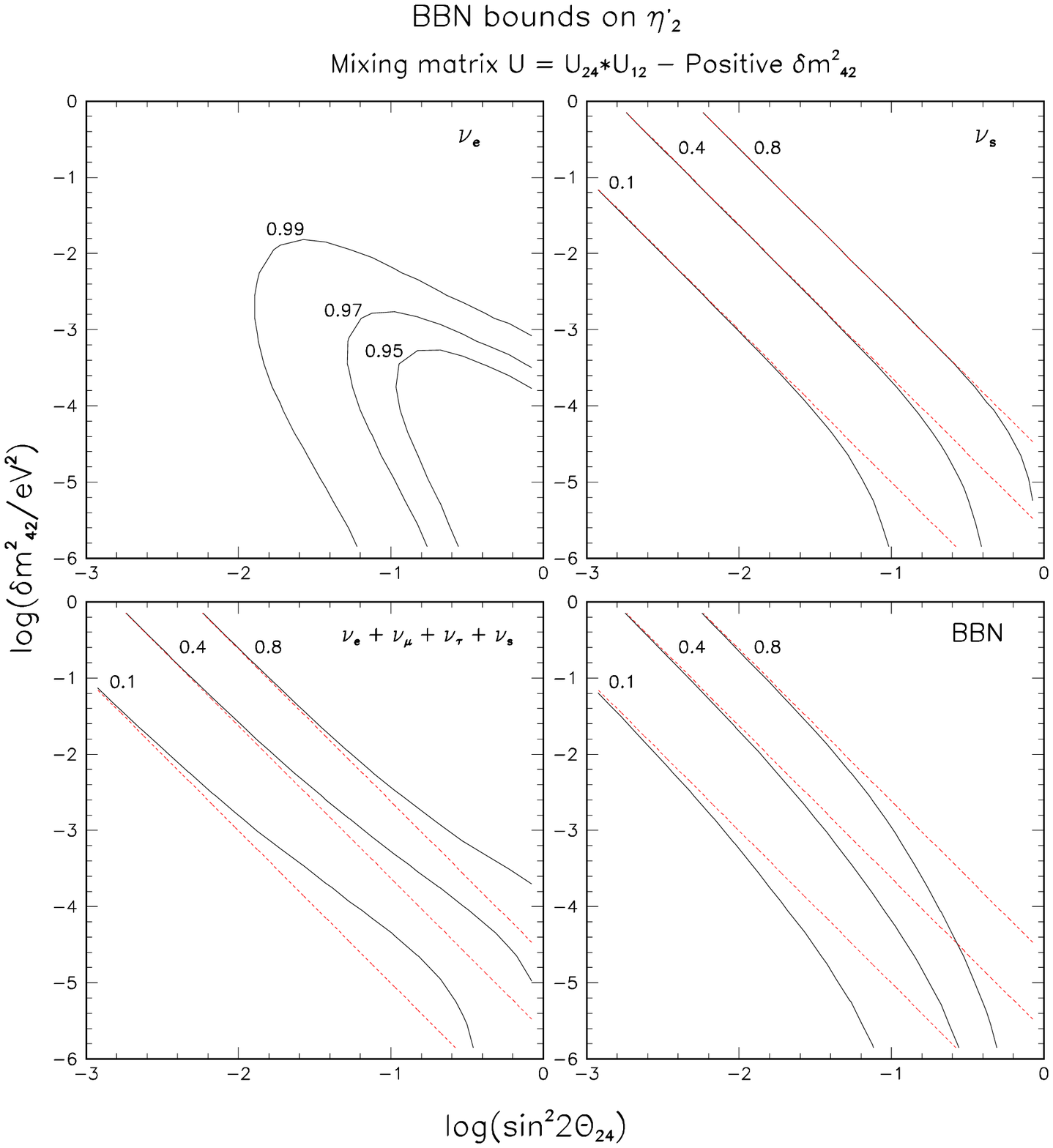,height=18cm,width=15cm}
\end{center}
\bigskip
\caption{Numerical results for the case in which the neutrino 
mixing matrix can be written as $U = U_{24}\cdot U_{12}$, with
$\dm_{21}$ and $\theta_{12}$ fixed according to eq.~(\ref{lma}).
We consider positive values for the mass difference 
$\dm_{42}=m_{4}^2-m_{2}^{2}$.
See Fig.1 for a detailed explanations of the various panels and 
for a definition of the various lines. Red dotted lines correspond 
to analytic estimates obtained from eq.(\ref{eta2-pos}).
Note that in the small mixing angle limits, the angle $\theta_{24}$
coincides with the parameter $\eta'_{2}$ defined in eq.(\ref{eta2})}
\label{fig7}
\end{figure}

\begin{figure}[hbt]
\begin{center}
\epsfig{file=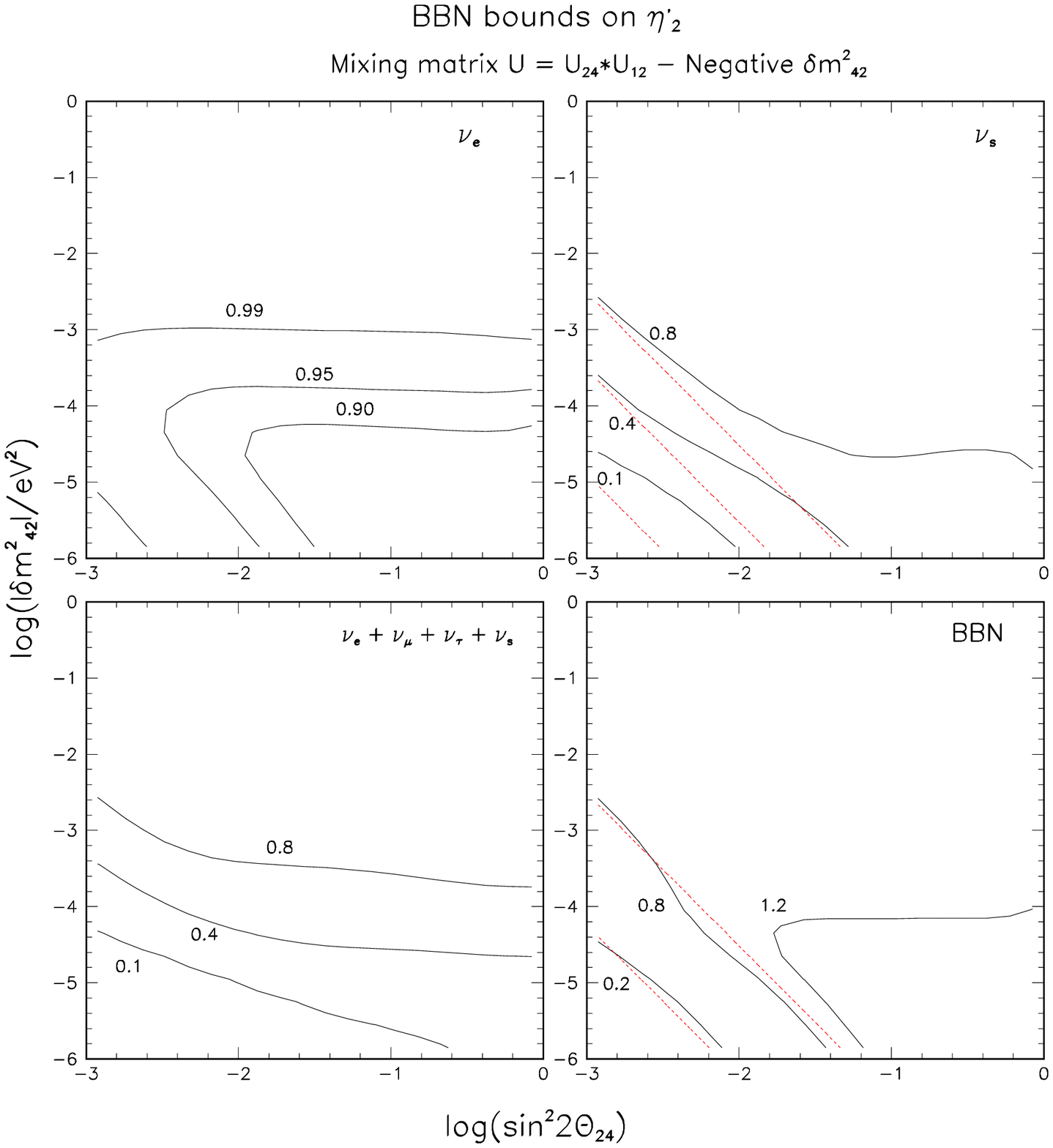,height=18cm,width=15cm}
\end{center}
\bigskip
\caption{Numerical results for the case in which the neutrino 
mixing matrix can be written as $U = U_{24}\cdot U_{12}$, with
$\dm_{21}$ and $\theta_{12}$ fixed according to eq.~(\ref{lma}).
We consider negative values for the mass difference 
$\dm_{42}=m_{4}^2-m_{2}^{2}$.
See Fig.1 for detailed explanation of the various panels and 
for a definition of the various lines. Red dotted lines correspond 
to analytic estimates obtained from eq.(\ref{eta2-neg}).
Note that in the small mixing angle limits, the angle $\theta_{24}$
coincides with the parameter $\eta'_{2}$ defined in eq.(\ref{eta2})}
\label{fig7bis}
\end{figure}

\begin{figure}[hbt]
\begin{center}
\epsfig{file=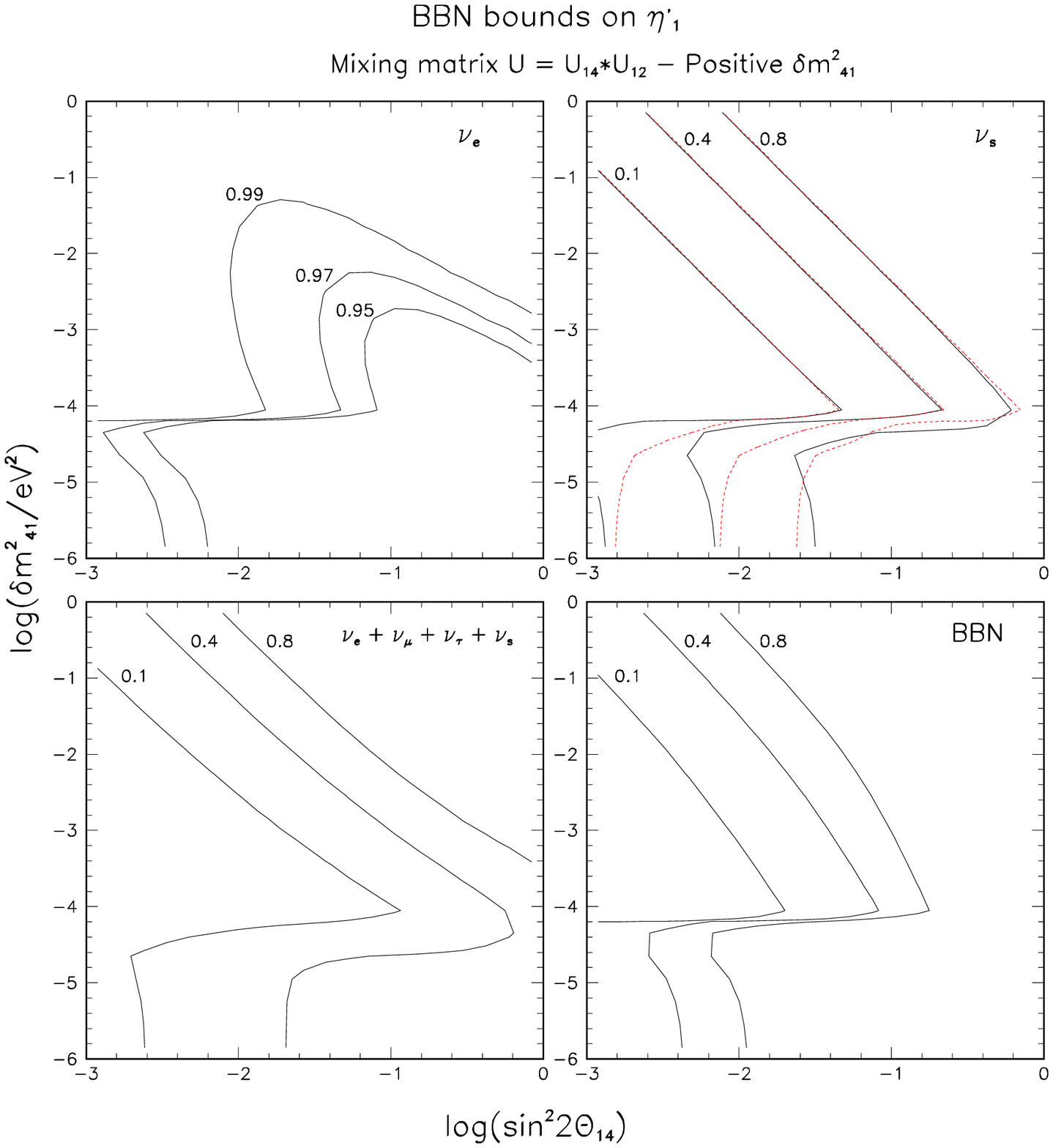,height=18cm,width=15cm}
\end{center}
\bigskip
\caption{Numerical results for the case in which the neutrino 
mixing matrix can be written as $U = U_{14}\cdot U_{12}$, with
$\dm_{21}$ and $\theta_{12}$ fixed according to eq.~(\ref{lma}).
We consider positive values for the mass difference 
$\dm_{41} = m_{4}^2-m_{1}^{2}$.
See Fig.1 for detailed explanation of the various panels and 
for a definition of the various lines. Red dotted lines correspond 
to analytic estimates obtained from eqs.(\ref{eta1-pos},\ref{phi-neg}).
Note that in the small mixing angle limits, the angle $\theta_{14}$
coincides with the parameter $\eta'_{1}$ defined in eq.(\ref{eta1})}
\label{fig8}
\end{figure}

\begin{figure}[hbt]
\begin{center}
\epsfig{file=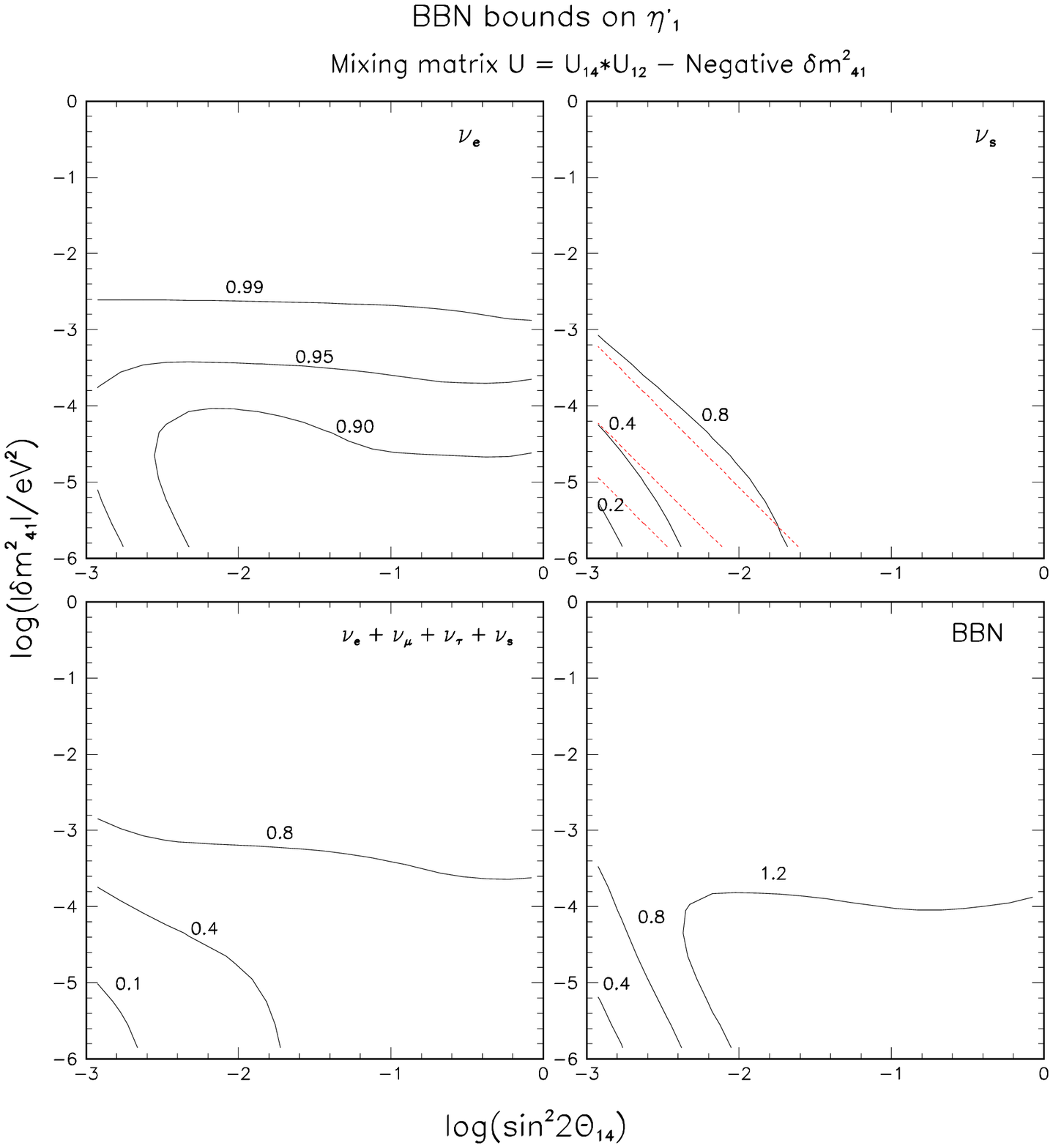,height=18cm,width=15cm}
\end{center}
\bigskip
\caption{Numerical results for the case in which the neutrino 
mixing matrix can be written as $U = U_{14}\cdot U_{12}$, with
$\dm_{21}$ and $\theta_{12}$ fixed according to eq.~(\ref{lma}).
We consider negative values for the mass difference 
$\dm_{41} = m_{4}^2-m_{1}^{2}$.
See Fig.1 for detailed explanation of the various panels and 
for a definition of the various lines. Red dotted lines correspond 
to analytic estimates obtained from eq.(\ref{eta1-neg}).
Note that in the small mixing angle limits, the angle $\theta_{14}$
coincides with the parameter $\eta'_{1}$ defined in eq.(\ref{eta1})}
\label{fig8bis}
\end{figure}

We have shown above that the effects of strong mixing between all
 three active neutrinos
noticeably change the BBN bounds on the possible
 admixture of sterile neutrinos previously obtained under
 assumption of no active-active mixing. 
Our results are summarized in sec. 6 with simple analytical expressions.
They essentially depend upon the sign of mass difference between different 
neutrino species, because the latter determines the number and positions of the 
resonances. More detailed and more accurate bounds can be read-off from figures 
\ref{fig1},\ref{fig5},\ref{fig7},\ref{fig7bis},\ref{fig8},\ref{fig8bis},
 panel 4 in all the cases. The results depend upon the limits on the number of 
effective neutrino flavors allowed by BBN. One can see
that even with very modest limits there is not much freedom permitted to 
admixture of sterile neutrinos to the known active ones.

If lepton asymmetry of the universe is non-negligible the conclusions are less 
definite because of the additional parameter, the value of the asymmetry. 
However, if  asymmetry is generated by the oscillations themselves, the mixing
 with $\nus$ should be very small, below the bounds derived for the case of 
vanishing asymmetry. 
For the case of primordial asymmetry the bounds are weaker and with some 
conspiracy they could be very weak, allowing rather strong mixing between 
sterile and active neutrinos. 
More detailed investigation is desirable in this case.

 \bigskip
 
{\bf Acknowledgment} A.D. Dolgov is grateful to the Research Center for
the Early Universe of the University of Tokyo for the hospitality during
the time when this work was completed. 
F.L. Villante is grateful to A.~Mirizzi, D.~Montanino, M.~Giannotti and 
D.~Comelli for useful suggestions and comments.



\end{document}